\documentclass[final,3p,times]{elsarticle}

\usepackage{amssymb}
\usepackage{amsmath}

\usepackage{xcolor}
\usepackage{booktabs}
\usepackage{tabularx}
\usepackage{multirow}
\usepackage{float}
\usepackage[hyphens]{url}

\begin{document}

\begin{frontmatter}

\title{Engineering the Law-Machine Learning Translation Problem: Developing Legally Aligned Models} 

\author[1,3]{Mathias Hanson} 
\author[2,3]{Gregory Lewkowicz}
\author[1]{Sam Verboven}

\affiliation[1]{organization={Vrije Universiteit Brussel, Data Lab},
            addressline={Pleinlaan 2}, 
            city={Brussels},
            postcode={1050}, 
            state={},
            country={Belgium}}

\affiliation[2]{organization={Université Libre de Bruxelles - Smart Law Hub},
            addressline={Avenue Franklin Roosevelt 50}, 
            city={Brussels},
            postcode={1050}, 
            state={},
            country={Belgium}}

\affiliation[3]{organization={FARI - AI for the Common Good Institute},
            addressline={Cantersteen 10/12}, 
            city={Brussels},
            postcode={1000}, 
            state={},
            country={Belgium}}

\begin{abstract}
Organizations developing machine learning-based (ML) technologies face the complex challenge of achieving high predictive performance while respecting the law. This intersection between ML and the law creates new complexities. As ML model behavior is inferred from training data, legal obligations cannot be operationalized in source code directly. Rather, legal obligations require "indirect" operationalization. However, choosing context-appropriate operationalizations presents two compounding challenges: (1) laws often permit multiple valid operationalizations for a given legal obligation\textemdash each with varying degrees of legal adequacy; and, (2) each operationalization creates unpredictable trade-offs among the different legal obligations and with predictive performance. Evaluating these trade-offs requires metrics (or heuristics), which are in turn difficult to validate against legal obligations. Current methodologies fail to fully address these interwoven challenges as they either focus on legal compliance for traditional software or on ML model development without adequately considering legal complexities. In response, we introduce a five-stage interdisciplinary framework that integrates legal and ML-technical analysis during ML model development. This framework facilitates designing ML models in a legally aligned way and identifying high-performing models that are legally justifiable. Legal reasoning guides choices for operationalizations and evaluation metrics, while ML experts ensure technical feasibility, performance optimization and an accurate interpretation of metric values. This framework bridges the gap between more conceptual analysis of law and ML models' need for deterministic specifications. We illustrate its application using a case study in the context of anti-money laundering.
\end{abstract}

\begin{keyword}
artificial intelligence \sep machine learning \sep regulatory compliance \sep legal risk management \sep AI governance \sep human-centric AI \sep anti-money laundering
\end{keyword}

\end{frontmatter}

\section{Introduction}

In recent years, governmental institutions and private organizations have been rapidly adopting machine learning (ML) technologies in their operations for a variety of applications, such as, tax audits \citep{battaglini2024refining, de2018tax} anti-money laundering monitoring \citep{chen2018machine}, customer churn prediction \citep{rahman2020machine, de2024hybrid}, and clinical and operational decision-making in healthcare \citep{verboven2021combining}. This widespread adoption is driven by the promise of superior predictive performance, which in turn claims to enhance the associated decision-making.

Concurrently, these organizations must comply with all applicable laws and demonstrate \textit{ex ante} consideration for the law at the ML model design stage (as foreseen in, \textit{e.g.}, the EU AI Act \citep{malgieri2024licensing}). However, the law is generally abstract and open for different interpretations \citep{asgeirsson2020nature, hildebrandt2021adaptive, weerts2023algorithmic}. The question if a certain interpretation of a legal obligation and the application thereof is legally valid is inherently uncertain \textit{ex ante} a final and binding judgment by a controlling authority \citep{dari2007uncertainty, siena2009designing, otto2007addressing}. Regardless, organizations must still choose a course of action when designing their operations. As they can only make "estimates of the likelihood of liability attached to each course of action" before a judgment \citep{calfee1984some}, they must thus make a best effort attempt to ensure that the introduction of the ML model does not lead to a violation of the law---assuming this is the desired outcome\footnote{As mentioned in \citep{calfee1984some}, "liability-conscious [organisations may] consider the probability that they will or will not be held liable [or be sanctioned financially or otherwise] at a greater or lesser level of compliance."}---and document their choices to reflect how they considered these aspects at the ML model design stage \citep{malgieri2024licensing}.

Such a best effort attempt generally requires (1) the "operationalization" of (certain) legal obligations and (2) evaluating ML models' impact on each legal obligation. In this context, "operationalizing" a legal obligation refers to the conversion of abstract requirements stemming from the law into specific, implementable measures that can be integrated into the ML model design\textemdash which will often also be reductive. For instance, the prohibition to discriminate on the basis of certain characteristics could be operationalized through an algorithm that optimizes, depending on the context \citep{weerts2023algorithmic}, for demographic parity or equalized odds \citep{hardt2016equality}. The legal principle of data minimization could be operationalized through a combination of a performance-based criterion \citep{shanmugam2022learning, goldsteen2022data} and a k-anonymization technique (see, \textit{e.g.}, \citep{goldsteen2023ai}) and product liability laws may require operationalization through the use of inherently interpretable models \citep{rudin2019stop, fresz2024should}. To evaluate the impact on the different legal obligations, metrics on the ML model's performance on different ethical aspects (\textit{e.g.} "fairness" and "privacy") are recognized as informative in the literature. For instance, \citet{WACHTER2021105567} propose the metric "conditional demographic disparity" to aid in assessing automated discrimination cases. The data minimization principle could be evaluated on the basis of the percentage of data used and the re-identification risk calculated using population uniqueness after applying k-anonymity \citep{bandara2020evaluation}. The literature on ML and ethical concepts (such as fairness, explainability, and privacy) thus serves as a crucial resource, providing valuable inspiration for both operationalizations and evaluation metrics. 

At the same time, this literature also reveals a fundamental challenge of trade-offs when considering different ethical concepts simultaneously. For instance, it has been observed that improvement in one area (\textit{e.g.}, privacy) can lead to degradation in another area (\textit{e.g.}, fairness) \citep{agarwal2021tradepriv, gu2022privacy, sun2023tradeoffs, ferry2023sok}. Furthermore, each operationalization may also have an \textit{a priori} unknown adverse effect on predictive performance. As operationalizations of the law are based on this body of literature, similar trade-offs will arise in legal contexts. For the "right to be forgotten" and the "right to an explanation", trade-offs are even empirically confirmed \citep{pawelczyk2023tradeoff}.

Despite the overlap between law and ethics \citep{xhemajli2021role, gardner201035, rentmeester2006s, lewkowicz2024technical}, core differences between both fields lead to unique challenges when dealing with trade-offs at the ML model level. In the field of ethics, one can argue that a specific metric represents an ethical concept and that achieving a certain value on this metric means this ethical concept has been successfully implemented. For instance, a model has been considered "fair" when its training algorithm is optimized for \textit{e.g.}, demographic parity or equalized odds \citep{hardt2016equality}\footnote{\label{note1}This approach has, however, been criticized by demonstrating the limitations of ML models optimized for fairness to actually achieve "fair" outcomes for the real world, see, \textit{e.g.}, \citep{selbst2019fairness}.}. The nature of legal obligations, on the other hand, prevents such an individually arguable standard\footnote{To avoid confusion, we do not argue that such an individually arguable standard means that "ethics is subjective". For a discussion on this, see \citep{blackman2022ethical}.}. Instead, lawyers attempt to "estimate" what controlling authorities would consider the legal obligation to be. This means that the standard that should be met is inherently uncertain. At the same time, legal obligations do not map directly to an operationalization or a certain value on a given metric as these methods do not fully capture what is required by legal obligations. For instance, it cannot be assumed that no illegal discrimination has occurred merely because a training algorithm was optimized for a specific fairness metric \citep{WACHTER2021105567, weerts2023algorithmic, koumeri2023compatibility}. Rather, it is generally possible to construct valid legal arguments defending different methods of operationalization and evaluation for a specific context.

Organisations thus face particularly complex challenges when developing, evaluating and selecting ML models, especially in highly regulated industries (\textit{e.g.,} finance \citep{kerrigan2022artificial, heavily-regulated-industry} and health care \citep{shen2023ai}). They need to satisfy multiple legal mandates---without clear indications on whether these are actually satisfied---while achieving high predictive performance for improved decision-making. Current methods fall into two approaches, both failing to adequately address this challenge. Some methods focus on developing legally compliant "traditional software", where legal evaluation can be performed by reviewing the code line by line (see, \textit{e.g.}, \citep{hassani2024rethinking, sleimi2019query, akhigbe2019systematic, abualhaija2023legal, fassbender2017domain, hoess2024towards}). In ML models, legal alignment can only be assessed "indirectly" through evaluation metrics or heuristics. Other methods focus on developing ML models without adequately considering the complexities introduced by requirements stemming from the law (see, \textit{e.g.}, \citep{gjorgjevikj2023requirements, zowghi2023s, budiardjo2024roadmap, ahmad2023requirements, martinez2022software, de2023classification}). Specifically, there are no methods related to how the ambiguity of laws and the complex relations between statements within a single law and between diﬀerent laws \citep{fassbender2017domain, boella2014critical} can be addressed in the context of ML model development.

In this paper we propose a novel framework for developing, evaluating and selecting ML models when different legal obligations apply simultaneously. Our framework facilitates legal justification, and provides insight into the impact of different operationalizations of legal obligations on trade-offs at the ML model level between proxy metrics for legal obligations, and predictive performance. Using our framework, organizations can: (1) develop ML models with due consideration for legal obligations; (2) evaluate the impact of different legal operationalizations on each applicable legal obligation and ML performance; and, (3) make informed model selection decisions based on legally substantiated justifications and predictive performance. By providing a structured approach to navigate the complex landscape of legal obligations and their technical implementations, our framework empowers organizations to develop and deploy ML models that optimizes predictive performance within the---likely---boundaries of the applicable legal framework.

\section{Related Works}

\label{Literature_review}

Current literature does not address the development, evaluation and selection of ML models for a real-world context where multiple legal obligations apply simultaneously. Nevertheless, there exists relevant software engineering (SE)---including requirements engineering (RE)---literature. SE is defined as "the application of a systematic, disciplined, quantifiable approach to the development, operation, and maintenance of software" \citep{159342, van2008software}. The field of RE, which is considered a crucial phase in the SE lifecycle, is "concerned with specifying and maintaining requirements of a system-to-be" \citep{abualhaija2023legal}. In both these fields, there has been increasing attention for developing software (requirements) to comply with the law, as well as for developing approaches tailored to ML software systems \citep{ahmad2023requirements, martinez2022software}.

However, SE---including RE---literature focusing on requirements stemming from the law is not adjusted for the challenges posed by the data-driven nature of ML models \citep{gjorgjevikj2023requirements, zowghi2023s, budiardjo2024roadmap, ahmad2023requirements}. Unlike “traditional software” systems where requirements can be assessed by reviewing code line by line, ML models present unique challenges as they must be evaluated indirectly through metrics and heuristics \citep{belani2019requirements, ahmad2023requirements}. At the same time, SE---including RE---literature that focuses on ML model-based software (SE4AI and RE4AI) falls short for the challenges of requirements emanating from legal obligations. Such law-based requirements differ from other types of requirements as they originate from a complex system of binding rules and norms that interact and are enforced by a controlling authority \citep{hildebrandt2021law, mw:law}, creating unique challenges for designing requirements for ML models or ML model-based software (see, \textit{e.g.}, \citep{boella2014critical, fassbender2017domain}), including the navigation of inherent uncertainty and the need for a holistic legal evaluation---as set out under section \ref{Need_new_method}

\subsection{Current Lawful Software and Requirements Engineering} The goal of requirement engineers when dealing with legal obligations is four-fold \citep{otto2007addressing, ghanavati2016making}: identify relevant legal provisions (see, \textit{e.g.}, \citep{hassani2024rethinking}), extract legal requirements and key concepts from the text (see, \textit{e.g.}, \citep{boella2014critical, sleimi2019query}), model the legal texts using goal-oriented RE or other formal notations (see, \textit{e.g.}, \citep{otto2007addressing, akhigbe2019systematic}), and monitor compliance through "machine-analyzable representations of legal text based on which automated analysis technologies can be developed" (see, \textit{e.g.}, \citep{abualhaija2023legal, castellanos2022compliance}). Unresolved issues or complexities in achieving these goals have been emphasized and attempts to address (some of) these have been proposed for "traditional software" \citep{otto2007addressing, boella2014critical, fassbender2017domain, hoess2024towards}. However, these methods fall short for software which include ML models.

In traditional software systems, an implementation of legal obligations can be directly embedded into the code, allowing for more straightforward legal evaluation through direct inspection of the code. ML models require a different method as these are fundamentally data-driven. Instead of following fixed, predetermined rules, the behavior of ML models is inferred from training data \citep{martinez2022software}. As an ML model's behavior thus emerges from processing training data rather than following explicit rules, legal requirements cannot be directly embedded in the training algorithm's code. Furthermore, assessing legal requirements by reviewing code line by line is impossible, as the model's behavior is determined by its learned parameters rather than the code itself \citep{belani2019requirements, ahmad2023requirements}. Rather, evaluation metrics or heuristics are the core tools for informing legal evaluation, making this evaluation more "indirect" \citep{belani2019requirements, ahmad2023requirements}. As a result, current methods for developing lawful or legally compliant "traditional software" are inadequate for ML model-based software.

\subsection{Current SE4AI / RE4AI Methods} The subfields of SE4AI / RE4AI emerged recently as the conventional methods for building, operating and maintaining traditional software were insufficient for addressing the challenges posed by ML model-based software \citep{gjorgjevikj2023requirements, zowghi2023s, budiardjo2024roadmap, ahmad2023requirements}. Identified challenges include defining data requirements, addressing new non-functional requirements related to ethics, trust, explainability and transparency \citep{ahmad2023requirementsEmpirical, ahmad2023requirementsFramework, ahmad2023requirementsElicitation}\footnote{\label{note2}Requirements for software are generally categorized as functional or non-functional requirements. In this paper, the term "legal requirement" refers to any requirement for software that originates from a legally binding source. Such requirements may be functional or non-functional depending on the underlying legal obligation(s).}, and navigating \textit{a priori} unkown trade-offs between different objectives like accuracy and fairness \citep{ahmad2023requirements, budiardjo2024roadmap, zowghi2023s, de2023classification}. In response to these challenges, recent literature has proposed various approaches to adapt SE and RE practices for ML model-based software. For instance, efforts have been made to address non-functional requirements related to diversity \citep{bano2023ai}, explainability \citep{sporsem2024discovering, balasubramaniam2024candidate}, and, ethics and trust \citep{damirchi2023non} in AI development. Additionally, researchers have advanced holistic processes and ontologies to better manage the complexities of ML model development \citep{martinsrequirements, sadovski2024towards}. However, how to address the challenges that arise when developing ML models where requirements stem from the law remains an open question. In the next section, we show why such requirements necessitate a new methodology.

\section{The Imperative for a New Methodology for Legal Obligations}

\label{Need_new_method}

Law is a complex system of binding rules and norms that interact and are prescribed, formally recognized as binding or enforced by a controlling authority \citep{hildebrandt2021law, mw:law}. Organizations must ensure that their ML models are aligned with all applicable legal obligations simultaneously. Failure to comply with any of them can result in sanctions. Consequently, developing ML models with respect for applicable legal obligations generally requires operationalizing applicable obligations and evaluating the ML model's impact on these obligations through an indicative proxy metric (or heuristic).

As operationalizing legal obligations and the ways of measuring trade-offs at the model level when implementing legal obligations often involves (adapted) tools from the ethics and ML literature \citep{weerts2023algorithmic, WACHTER2021105567, koumeri2023compatibility}, a logical extension is that similar \textit{a priori} unkown trade-offs will also emerge in case legal obligations apply. These trade-offs are demonstrated empirically in a body of work that pursues simultaneous optimization of two or more ethical aspects. In \citep{agarwal2021tradeinterpret}, \citep{jabbari2020empirical} and \citep{jo2022learning}, it is shown that there often is a trade-off between fairness and interpretability if accuracy is to be maintained. The trade-off between fairness, privacy and accuracy is also demonstrated in theory \citep{agarwal2021tradepriv}, and in practice \citep{pannekoek2021investigating, gu2022privacy, sun2023tradeoffs}. When jointly considering fairness, interpretability and privacy requirements, it is also difficult to preserve a high level of performance \citep{ferry2023sok}. Similarly, \citet{gittens2022adversarial} also demonstrate the multilateral trade-offs among accuracy, robustness, fairness and privacy. While  sometimes, trade-offs may be negligible \citep{rodolfa2021empirical}, it becomes increasingly challenging to jointly optimize without compromise as more constraints are considered simultaneously. Although the same types of \textit{a priori} unknown trade-offs emerge when implementing legal obligations and ethical concepts in ML models, fundamental differences between ethics and law make developing legally aligned ML models a distinctly different challenge.

\subsection{Generally Applicable Uncertain Legal Standards v. Individually Arguable Ethical Standards}
Law is binding and enforced by an external controlling authority---whether a court of law or another regulatory body. This is a crucial difference with ethics, in which there are no such binding standards, nor external authorities. As a result, in ethics, ML developers can argue their own standard to be met for ML model adherence to a certain ethical concept. For instance, some ML practitioners may consider an ML model to achieve "fair" results if it achieves a certain value on the \textit{e.g.} demographic parity or equalized odds \citep{hardt2016equality} metric. However, for legal compliance, organizations must argue on what the external authority would consider a "correct" interpretation and application of the law in a given context. Whether an organization has fully respected the law is thus inherently uncertain \textit{ex ante} a final and binding decision \citep{dari2007uncertainty, siena2009designing, otto2007addressing} by the external authority.

\subsection{Holistic Legal Assessment v. Ethical Assessment}
Legal assessment is concerned with evaluating if the obligations and rights the law imposes and grants have been respected. These obligations and rights are generally abstract and do not provide detailed instructions as to how these should be implemented in a specific context or for a specific technology \citep{ghanavati2016making}. Organizations must therefore rely on legally grounded choices when implementing the law. For ML models, this includes making legally substantiated choices for operationalizations and a legally grounded stance for which values on proxy metrics may be acceptable in a given context. However, while any operationalization or a value on a given metric may demonstrate plausibility that the introduction of the ML model in a broader system does not lead to a violation of legal obligations, it cannot show no legal obligations have been violated. For instance, the quantitative aspect of the legal principle of data minimization can be operationalized through a performance-based criterion that limits data collection by iteratively estimating the relationship between dataset size and system performance to establish a data acquisition stopping point \citep{shanmugam2022learning, goldsteen2022data}. However, relying on this operationalization does not guarantee the legal principle has been respected. Data minimization requires limiting the quantity of personal data used and "anonymizing or pseudonymizing wherever possible" given a specific purpose \citep{https://doi.org/10.26116/techreg.2021.004}. Whether the principle has been respected thus depends on several factors, such as, the suitability of the chosen operationalization for limiting data quantity within the specific context, the threshold used to establish the stopping point for data acquisition, the extent to which the data has been anonymized and pseudomyzed, and what the specified purpose is. In essence, a legal assessment requires giving due weight to the chosen operationalizations and the achieved values on relevant metrics, but ultimately focuses on the the ML model's alignment with applicable legal obligations and rights \citep{WACHTER2021105567, weerts2023algorithmic, koumeri2023compatibility}.

In contrast, in the field of ethics, a certain operationalization or a value on a given metric can be argued to be a sufficient representation of a certain ethical concept. For instance, there is a wide variety of operationalizations and metrics that are argued to capture the concept of "fairness" \citep{weerts2023algorithmic, pessach2022review}\footnote{See footnote \ref{note1}.}. Similarly, for the above-mentioned example, one could argue that the mentioned data minimization operationalization \citep{shanmugam2022learning, goldsteen2022data} is sufficient for meeting "privacy" requirements.

\subsection{The Consequences for Developing ML Models under Legal Obligations}
Developing legally aligned ML models means respecting all legal obligations simultaneously, while it is inherently uncertain how to achieve this. A good faith and best effort attempt means making legally justified choices so that any operationalization or achieved values on any metric aligns, to the extent possible, with the estimated "correct" interpretation and application of the applicable legal obligations. This requires legal reasoning, which  involves interpreting various authoritative legal sources, such as legislation, case law, and regulatory guidance, and reconciling these interpretations in a manner legal practitioners believe authorities would deem "correct" \citep{vandevelde2018thinking, rentmeester2006s} in a given context \citep{WACHTER2021105567, weerts2023algorithmic, bueno2024challenges}. This legal reasoning process inevitably involves some level of subjectivity. Consequently, legal arguments can often be made in favor of different operationalizations and metrics, as well as on which values on these metrics are likely suitable to (reductively) represent a certain legal obligation in a given context.

This flexibility to argue in favor of different operationalizations creates a new challenge. The literature on ethical concepts and ML shows that changing how an ethical concept is operationalized can unpredictably affect the values measured on metrics that represent other ethical concepts and predictive performance. Similarly, in the legal domain, adjusting the operationalization of a legal obligation alters the values on indicative proxy metrics that represent legal obligations and impacts predictive performance in ways that cannot be determined in advance. Because of this unpredictability, it is impossible at the development stage to guarantee that certain (legal) proxy metric values will be achieved under certain operationalizations. However, legal alignment of an ML model depends on both the chosen operationalizations and the resulting proxy metric values. Consequently, selecting an ML model in a legally regulated context, is not simply a matter of choosing the best performing model from a set of legally aligned models. Instead, any methodology for designing legally aligned, high-performing ML models must account for interactions between operationalizations, proxy metric values and predictive performance, and enable an understanding how these unfold at the model level.

Another new challenge is choosing or designing a legally meaningful metric. For that decision, the key is to understand why and when a certain metric will be most useful for the holistic legal evaluation \citep{WACHTER2021105567, weerts2023algorithmic}. This implies a thorough understanding of what any given metric precisely represents and recognizing that any metric has limitations for legal assessments. For instance, as metrics are aggregates, they inherently obscure individual contexts. Consequently, they may be useful for assessing system-level compliance, but they cannot capture all contextual elements that matter for protecting individual rights. A metric showing group-level non-discrimination could still mask individual instances of discrimination, which the legal system would recognize and seek to address through the individual right to non-discrimination \citep{hildebrandt2021law}.

In addition, well-designed metrics will also be time-limited for use in legal assessments due to the adaptive nature of legal norms \citep{hildebrandt2021adaptive, boella2014critical}. As noted by \citet{boella2014critical}, "[legal] [n]orms are living entities that emerge from a plurality of sources and adapt continuously, not only to legislative changes, but also to the way in which they are interpreted in different contexts by judges and legal scholars". As such, any metric for assisting in legal assessments at the time of model development may no longer be as meaningful when re-evaluating the model during deployment. These metrics need to continuously evolve in response to new legal rulings and interpretations.

\subsection{Conclusion: The Need for New Interdisciplinary Methodologies}

New interdisciplinary methodologies are thus essential for developing legally aligned ML models due to the combination of ML's data-driven nature, inherent legal uncertainty and the need for a holistic legal assessment. Legal reasoning must guide the selection of operationalizations and evaluation metrics or heuristics, but ML expertise is crucial for ensuring technical feasibility of operationalizations, the calculability of metrics and the usefulness of heuristics. In addition, ML experts must optimize model performance within these uncertain legal boundaries, despite \textit{a priori} unknown trade-offs at the model level. Lastly, for values achieved on certain metrics, it is crucial ML experts clearly communicate about what these values represent (and do not represent) so that legal experts can interpret these in a legally meaningful way. Without this integrated approach, engineers may adopt a simplified or naive interpretation of the law \citep{boella2014critical}, while lawyers may struggle to understand what operationalizations are technically feasible and how ML models may impact legal obligations. Only through interdisciplinary work can ML models be developed that meet what is legally required and achieve good predictive performance.

\section{Our contribution}
We propose an interdisciplinary framework to design, evaluate and select ML models. Using this framework, organizations garner insights into the trade-offs between different legal obligations at the model level and their impact on predictive performance under legal uncertainty. Furthermore, these insights are the basis for improved legal justification for model choice.

Our proposed framework involves the following five stages and is visually represented in Figure \ref{fig:visual}:

\begin{figure*}[t]
  \centering
  \includegraphics[width=\textwidth]{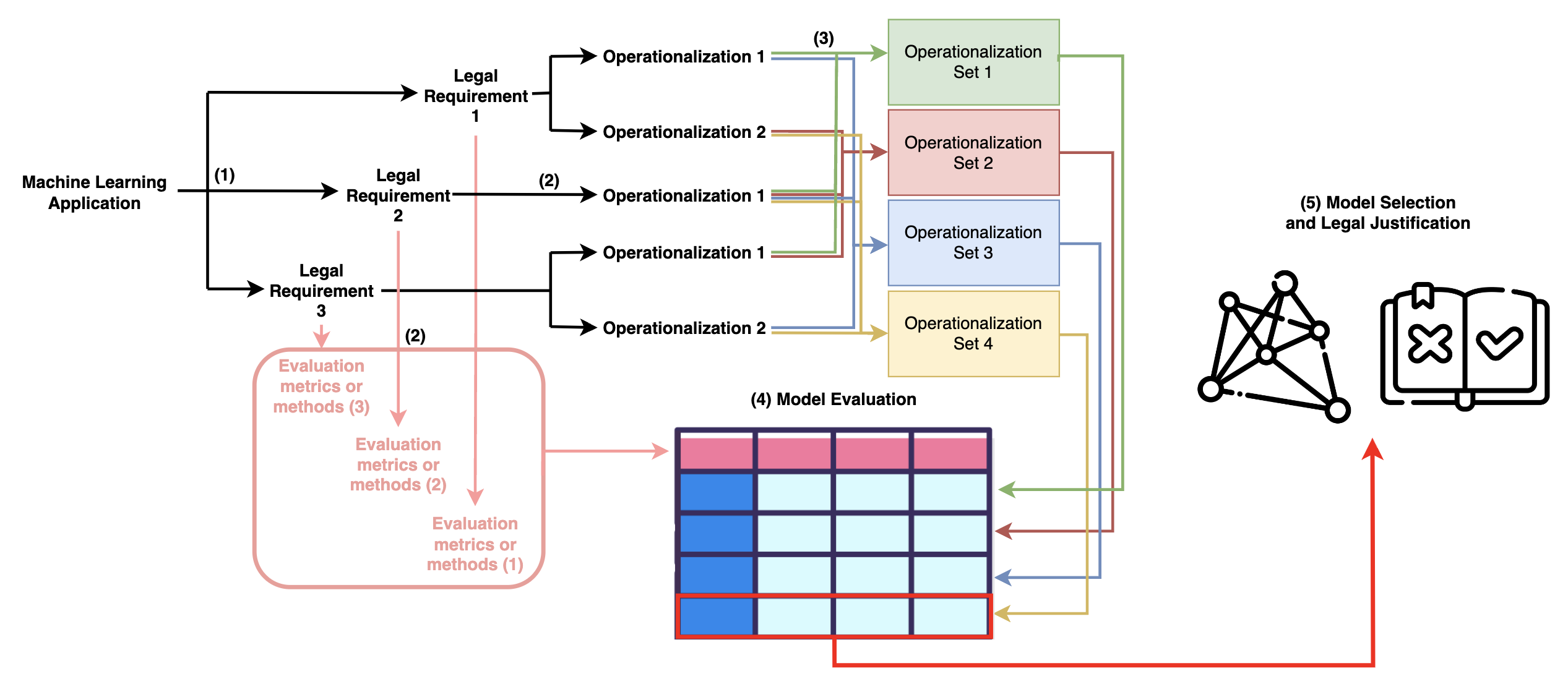}
  \caption{5-Stage Framework to Design, Evaluate and Select ML Models under Legal Obligations}
  \label{fig:visual}
\end{figure*}

\noindent \textbf{1. Identification of Legal Requirements}: Through a traditional legal analysis of all potentially relevant laws and regulations, the legal team first extracts the applicable legal obligations for the envisioned machine learning application. To avoid perverse incentives, the team takes a legally-grounded and substantiated stance on what they consider the "correct interpretation" to be of each legal obligation\footnote{Organizations must make a legally-grounded and substantiated estimate of which interpretation of a legal concept or obligation they believe authorities would deem "correct". For instance, they need to take a stance on whether the GDPR contains a right to explanation or not \citep{kaminski2020algorithmic, goodman2017european, selbst2018meaningful, malgieri2017right, bayamliouglu2022right, wachter2017right}. By taking a stance on such purely legal questions \citep{hoess2024towards}, organizations can set the "legal guardrails" within which their operations have to stay to comply with all applicable laws.}. Subsequently, the legal team work together with requirements engineers to translate these legal obligations to legal requirements\footnote{See footnote \ref{note2} on the term "legal requirement".} for the envisioned application \citep{hassani2024rethinking}, relying on existing methods for legal requirements elicitation (see, \textit{e.g.}, \citep{boella2014critical}). The output of Stage 1 is thus a detailed set of legal requirements to guide the subsequent technical operationalization and evaluation steps.

\noindent\textbf{2. Translation to Operationalizations and Evaluation Metrics (or Heuristics)}: A interdisciplinary team of legal and data science experts translates the legal requirements, where needed, into one or various technical operationalizations. For each legal requirement, an evaluation metric (or heuristic) that is most useful for the legal evaluation of the ML model is selected.
This approach is crucial to ensure that operationalizations are both legally grounded and technically feasible. It also guarantees that the evaluation metrics can be computed accurately and that the values achieved on these metrics can be evaluated in a legally meaningful way. Translating requirements into multiple operationalizations allows for the exploration of interactions between different legal obligations at the ML model level.

\noindent \textbf{3. Operationalization Set Formation and Model Training}: The data science team generates all feasible combinations of the identified operationalizations, ensuring technical compatibility. Each operationalization set comprises an operationalization of each legal requirement for which operationalizations were identified. If the number of operationalization sets is substantial, a subset may be selected based on anticipated legal alignment and predictive performance. Subsequently, the data science team runs the ML algorithms of choice (insofar as allowed by the legal requirements) on each selected subset to create the resulting set of models.

\noindent \textbf{4. Model Evaluation through Trade-off Mapping}: The data science team calculates the values on the quantitative metrics or describes the performance of each model using the relevant performance metrics and evaluation heuristics selected in Stage 2. The interdisciplinary team can then assess the results of each model across the different evaluated dimensions. This comparison facilitates the assessment of the impact of various operationalizations on performance metrics and proxy metrics for legal obligations. Given that enhanced performance as measured by a proxy metric for a specific legal obligations may lead to diminished performance in other evaluated dimensions, this process is referred to as "trade-off mapping". This is enabled by the groundwork laid in Stages 1-3, where legal obligations were thoroughly mapped out, translated to legal requirements, operationalized in multiple ways, and used to train a diverse set of models. 

\noindent \textbf{5. Model Selection and Legal Justification}: Based on the analysis conducted in Stage 4, the interdisciplinary team commits to deploying a model, fully informed about the trade-offs between the values on the indicative proxy metrics (and heuristics) for the legal obligations. The selected model optimizes legal alignment across all requirements, while (ideally) maintaining high predictive performance. The interdisciplinary team utilizes the results from Stage 4, grounded in metrics (and heuristics) from Stage 2, to develop a legal justification for the chosen model, making use, \textit{e.g.}, of a legal "proportionality" analysis \citep{custers2022new, tridimas2023wreaking, gellert2016we}. This argumentation substantiates the alignment of the final model with the legal obligations.

This framework enables organizations to convert legal obligations into operationalizations and evaluation metrics (or heuristics) for ML models. By assessing these metrics (or heuristics), organizations can understand trade-offs between different legal operationalizations and performance for each operationalization set. The trade-off mapping then supports building a substantiated legal argument for the model's legal alignment. While our framework does not exonerate organizations from taking mitigating measures for the risks that introducing the chosen model may bring for respecting the right of every single individual or for legal risks that are not represented by the evaluation metrics (or heuristics), it is a vital step for informed decision-making, accountability, responsible AI practices \citep{lin2019artificial} and for the mentioned risk mitigation. 

\section{Case study}
In what follows, we illustrate our proposed framework through a toy case study within the context of Anti-Money Laundering (AML) \citep{jensen2023fighting}. We consider the scenario where a financial institution develops a binary classifier to classify transactions as likely constituting money laundering or non-suspicious. Transactions predicted as positive indicate potential money laundering, while those predicted as negative are considered non-suspicious. A positive prediction by the model triggers a further investigation into the transaction for suspected money laundering activities. For this example, we only consider sources of European law, while recognizing that additional legal obligations may follow from national laws. The AI Act is not considered as it is unlikely to impose additional legal obligations for this anti-money laundering application that are relevant to our framework \citep{EU_AI_Act_2024, Parente_2024}.

We assume that a single observation in the dataset relates to one outgoing or incoming transaction to a bank account at our financial institution and would include values for following features and label:
\begin{enumerate}
    \item Personal information on the (main) account holder: Gender, Legal Domicile, Tax Residency, Source of Wealth Industry, Total Estimated Assets, Profession and Politically Exposed Person Status
    \item Transactional information: Direction (incoming or outgoing), Sender Account Number, Sender Country, Receiver Account Number, Receiver Country, Transaction Date, Transaction Type (securities or cash), Amount (in EUR) and Transaction Currency
    \item A binary label indicating whether a transaction constitutes likely money laundering or not. We assume that these labels represent the ground truth.
\end{enumerate}

\subsection{Stage 1: Identification of Legal Requirements}

The legal team identifies the applicable legal obligations that could be argued to apply to the envisioned ML application and takes a legally substantiated stance on what they consider the "correct interpretation" to be of each legal obligation in the given context. Subsequently, they, in cooperation with the requirements engineer(s), translate these legal obligations to legal requirements for the specific ML application.

We assume that the legal team conducts the following analysis:

\subsubsection{Non-Discrimination Requirements:} Given that the available data includes the feature "gender", a legal analysis is required to identify if there exists a legal basis that prohibits discrimination for the specific application in the case at hand. While EU anti-discrimination law is a broad field with many different sources of law between which the exact relationship is not always clear \citep{xenidis2019eu, EUNonDiscriminationLaw2018}, the consensus in the literature is that any discrimination on the basis of sex in "horizontal relationships" constitutes a direct violation of European law \citep{de2011novel, muir2019horizontal}. As such, the envisioned application should not illegally discriminate over sex by generating an alert merely due to membership of a certain "gender" category \citep{bertrand2021ai, sciurba2019incompatibility}.

\subsubsection{Data Protection Requirements:} European law introduces different potential legal obligations for the envisioned ML model in terms of data protection:

\textbf{Principle of proportionality}: In its case-law, the Court of Justice of the EU has established the principle that intrusive monitoring should be used only for the most serious crimes. \citet{bertrand2021ai}, argues that financial transaction monitoring systems violate this principle as they "cover a broad range of offenses going from terrorism to tax fraud". However, it is decided not to operationalize such proportionality in relation to different offenses as the ML application will only generate an alert for suspicions of the offense "money laundering". The model will thus not flag potential "terrorism financing" transactions, nor be trained on data identifying the crime that preceded money laundering (\textit{e.g.}, tax fraud). The level of intrusiviness of the monitoring by the ML model will thus be the same for all clients and transactions.

\textbf{Data minimization}: According to this principle under the European Union’s General Data Protection Regulation ("GDPR"), only personal data that is adequate, relevant and limited to what is necessary shall be processed \citep{GDPR2016, shanmugam2022learning}. As the envisioned ML model is developed using mainly personal data, respecting this principle implies limiting the quantity of personal data used and anonymizing or pseudonymizing wherever possible \citep{https://doi.org/10.26116/techreg.2021.004}.

\textbf{Right to contest automated decisions}: \citet{binns2021your} have considered that automated anti-money laundering detection systems could fall under the scope of article 22 of the GDPR, even if there is human intervention afterwards. This view seems to be supported by the CJEU's reasoning in the SCHUFA judgment \citep{CJEU_C634_21}. Article 22 would then provide clients of the financial institution the right to contest the output of the ML model in case of, \textit{e.g.}, the closing or freezing of his or her bank account. As the majority view in the literature is that contestability includes at least a limited right to explanation \citep{kaminski2020algorithmic, goodman2017european, selbst2018meaningful, malgieri2017right, bayamliouglu2022right, wachter2017right}, it is an open question which exact requirement this would introduce for ML models. However, this will not be operationalized in the ML model as the decision to, \textit{e.g.}, close or freeze a bank account will be made only on the basis of human judgment. By introducing measures to ensure there is \textit{meaningful} human intervention, the institution argues that the ML model falls out of the scope of article 22 of the GDPR.

\textbf{Data storage limitation, right to erasure, and resilience against re-identification attacks}: According to several authoritative sources \citep{SHRISHAK2025, veale2018algorithms, juliussen2023algorithms}, ML models may qualify as "personal data" under the GDPR if it is possible to infer information about individuals in the training data through model inversion or membership inference attacks. If this premise is accepted, 3 additional legal requirements have to be operationalized: (1) the capability to remove specific personal data from an ML model upon request (\textit{i.e.}, the right to erasure); (2) the capability to automatically remove personal data that is no longer necessary to retain (\textit{i.e.}, the data storage limitation principle); and, (3) resiliency to the re-identification attacks that could leak the training data. While \citet{leiser2020governing} have argued that ML models cannot qualify as personal data, the legal team does not consider this position tenable.

\subsubsection{Anti-money Laundering Requirements:} The various European legislative sources related to anti-money laundering (AML) require financial institutions to implement diverse measures, including the fulfillment of obligations commonly referred to as "know your customer" and "customer due diligence" \citep{king2018palgrave, ec_aml_cft}. To ensure financial institutions comply with these obligations, supervisory authorities conduct inspections or audits. Although there are no explicit legal obligations for ML models in this domain, financial institutions should demonstrate that their ML-based transaction monitoring system provides adequate risk coverage for money laundering activities in case of an audit. Therefore, the financial institution will only develop (1) ML models with a reasonable degree of (perceived) model explainability and (2) ML models that can demonstrate the preservation or enhancement of the current anti-money laundering risk coverage.\\

To summarize, we assume the team of legal and RE experts identifies the following five legal requirements for the ML model:

\begin{itemize}
    \item Data Minimization Requirement
    \item Non-discrimination Requirement
    \item Avoiding the Model as "Personal Data" Qualification Requirement
    \item AML Explainable Model Requirement
    \item AML Risk Coverage Requirement
\end{itemize}

\subsection{Stage 2: Translation to Operationalizations and Evaluation Metrics (or Heuristics)}

The interdisciplinary team comprising legal and data science experts translates each legal requirement, where needed, into possible technical operationalizations. For each legal requirement, an evaluation metric (or heuristic) is selected to aid in assessing if all legal obligations have been respected. 
In our case study, we assume that the interdisciplinary team decides to operationalize and evaluate each identified legal requirement as follows.

\subsubsection{Non-discrimination Requirement} 
\noindent \textbf{Operationalizations}: (1) The feature "gender" is deleted from the dataset before training as this should avoid direct discrimination (fairness through unawareness); and, (2) in addition to operationalization (1), the "reject-option classification" technique by \citet{kamiran2012decision} is applied to get more similar outputs over the different "gender" feature. While there are still many other options for operationalization both at the data and model level \citep{kenfack2023exploring, weerts2023algorithmic}, other options will not be explored to limit the amount of operationalization sets.

\noindent \textbf{Evaluation}: The non-discrimination requirement will be assessed using the concept of "conditional demographic disparity" over the "gender" feature, conditioned on the "source of wealth industry" and "total estimated assets" features. While other methods of evaluation could be argued according to \citet{wachter2020bias}, the same authors have proposed conditional demographic disparity as a standard baseline statistical measurement that aligns with the European Court of Justice "gold standard" for assessment of \textit{prima facie} discrimination \citep{WACHTER2021105567}.

\subsubsection{Data Minimization Requirement} 
\noindent \textbf{Operationalizations}: (1) To limit the amount of training instances, the Framework for Inhibiting Data Overcollection \citep{shanmugam2022learning}, a type of performance-based data minimization, will be applied with a low stopping threshold of -1.0e-07. Recognizing  other thresholds could be argued, this low threshold is considered appropriate because performance is crucial to meet legal obligations from the AML legislation; and, (2) in addition to operationalization (1), a k-anonymity algorithm (k = 7) will be applied to the training instances. This will involve generalizing the quasi-identifiers (personal information of the main account holder and transaction amount) before training the model. The chosen value for k is considered suitable \textit{a priori} as it is expected to suffice from a data protection perspective, while allowing high utility for the ML model.

\noindent \textbf{Evaluation}: Implementing the data minimization principle requires limiting the quantity of personal data used and anonymizing or pseudonymizing wherever possible \citep{https://doi.org/10.26116/techreg.2021.004}. As such, this requirement will be evaluated in two ways: (1) quantitatively in terms of \% of the available data that was used to train the model; and, (2) qualitatively where the application of k-anonimity will be considered as a pseudonymization technique.

\subsubsection{Avoiding the Model as "Personal Data" Qualification Requirement}
\noindent \textbf{Operationalizations}: (1) No operationalization is required. According to the risk-based approach enshrined in the GDPR, "organizations are required to calibrate the legal norms in the GDPR with an eye to the risks posed to the rights and freedoms of individuals" \citep{quelle2018enhancing}. Since the ML model will not be publicly shared or traded, any risk to the rights and freedoms of the data subject caused by potential re-identification is considered mostly theoretical. In addition, the team argues that the specific envisioned ML would not qualify as "personal data" relying on the work of \citet{shokri2017membership}. According to this work, a binary classification system trained on tabular data may not send sufficient signal for an attacker to extract useful membership inference from the model; and, (2) a k-anonimity algorithm (k = 7) will be applied on the quasi-identifiers (personal information on the (main) account holder and the transaction amount) to generalize these features before training the model. Many other defenses against membership inference attacks could be applied both at the data and model level \citep{jegorova2022survey}, but these will not be explored to limit the amount of operationalization sets. This second operationalization exceeds what the legal team estimates is required by law; however, given legal uncertainty, implementing such a prudent safeguard is warranted.

\noindent \textbf{Evaluation}: This requirement will be evaluated using five qualitative risk categories for re-identification: very low, low, moderate, high, and very high. Ideally, more rigorous methods, such as accuracy after membership inference attacks based on prediction labels \citep{yeom2018privacy, pmlr-v139-choquette-choo21a} or the effectiveness of model inversion attacks \citep{dibbo2023model}, would be used to empirically assess re-identification risks. However, the team deems the evaluation using risk categories acceptable as the model will not be publicly shared or traded.

\subsubsection{AML Explainable Model Requirement}
\noindent \textbf{Operationalizations:} (1) Logistic regression is chosen for its relative high explainability \citep{kobayashi2024explainable} to supervisory authorities; additionally, (2) random forest is implemented due to both moderate technical explainability \citep{kobayashi2024explainable} and strong perceived interpretability \citep{herm2023stop}. Both models are also considered to have strong predictive performance for the task at hand.

\noindent \textbf{Evaluation:} Logistic regression and random forest models will simply be assumed to be respectively "moderately" and "highly" explainable based on insights from the literature \citep{kobayashi2024explainable, herm2023stop}.

\subsubsection{AML Risk Coverage Requirement}
\noindent \textbf{Operationalizations:} Cost-sensitive learning is used with higher weights assigned to positive instances (\textit{i.e.}, money laundering cases) \citep{fernandez2018cost}. This increases the coverage of money laundering risks, aligning with the priorities of supervisory authorities. No other operationalizations are considered.

\noindent \textbf{Evaluation:} The performance on false negatives will be evaluated using recall.

\subsection{Stage 3: Operationalization Set Formation and Model Training}
Each operationalization set represents a unique combination of operationalizations for each legal requirement, ensuring that all relevant legal aspects are addressed. For our case study, we have identified five legal requirements. The operationalizations for each requirement are as follows:
\begin{itemize}
\item Non-discrimination Requirement: (1) Fairness through unawareness and (2) the reject-option classification in addition to operationalization (1)
\item Data minimization Requirement: (1) Framework for Inhibiting Data Overcollection with low stopping thresholds and (2) k-anonymity algorithm (k=7) in addition to operationalization (1)
\item Avoiding model as "personal data" qualification Requirement: (1) No operationalization, and (2) k-anonymity algorithm (k=7)
\item AML Explainable Model Requirement: (1) Logistic regression models and (2) Random Forest models
\item AML Risk Coverage Requirement: Assigning higher weights for positive instances
\end{itemize}

Operationalization (1) of "avoiding model as personal data qualification" cannot be combined with operationalization (2) of "data minimization". This is because applying the k-anonymity algorithm in operationalization (2) of "data minimization" automatically implies the use of operationalization (2) of "avoiding model as personal data qualification". As such, we arrive at 8 operationalization sets as shown in Table \ref{tab:operationalization_sets}.

\begin{table*}[htbp]
\caption{Operationalization sets for the case study}
\label{tab:operationalization_sets}
\small
\begin{tabular}{|p{0.35\linewidth}|p{0.05\linewidth}p{0.05\linewidth}p{0.05\linewidth}p{0.05\linewidth}p{0.05\linewidth}p{0.05\linewidth}p{0.05\linewidth}p{0.05\linewidth}|}
\hline
\textbf{Legal Requirement} & \textbf{Set 1} & \textbf{Set 2} & \textbf{Set 3} & \textbf{Set 4} & \textbf{Set 5} & \textbf{Set 6} & \textbf{Set 7} & \textbf{Set 8} \\ \hline
\textbf{Non-discrimination} & (1) & (1) & (1) & (1) & (2) & (2) & (2) & (2) \\ \hline
\textbf{Data minimization} & (1) & (1) & (2) & (2) & (1) & (1) & (2) & (2) \\ \hline
\textbf{Personal data qualification} & (1) & (1) & (2) & (2) & (1) & (1) & (2) & (2) \\ \hline
\textbf{AML Explainable Model} & (1) & (2) & (1) & (2) & (1) & (2) & (1) & (2) \\ \hline
\textbf{AML Risk Coverage} & (1) & (1) & (1) & (1) & (1) & (1) & (1) & (1) \\ \hline
\end{tabular}
\end{table*}

These operationalization sets serve as the configurations for training ML models. By training models for each set, the data science team ensures that all identified legal requirements are operationalized in a certain way. At this stage, it may become apparent that implementing certain combinations of operationalizations may not be valuable to pursue or practically unfeasible from a technical perspective. If such issues occur, Stage 2 and the "operationalization set formation" may need to be revisited.

\subsection{Stage 4: Model Evaluation through Trade-off Mapping}
The data science team evaluates each trained model based on two key dimensions: performance on the core ML task and alignment with each identified legal requirement (using the evaluation proxy metrics (or heuristics) defined in Stage 2). This comprehensive evaluation provides insight into how different operationalizations of legal requirements interact and their impact on predictive performance, allowing a team of legal experts and data scientists to assess the results of each model in a holistic manner.

On the basis of insights from \citep{agarwal2021tradeinterpret, agarwal2021tradepriv, jabbari2020empirical, ferry2023sok, gittens2022adversarial}, Table \ref{tab:evaluation_dimensions} was generated. It contains an overview of possible results for all evaluation criteria for each trained model.

\begin{table*}[htbp]
\caption{Evaluation dimensions for the case study}
\label{tab:evaluation_dimensions}
\resizebox{16.5cm}{!}{%
\begin{tabular}{l|ccc|cc|c|c|cc}
\toprule
\textbf{Number Set} & \multicolumn{3}{c}{\textbf{Predictive Performance}} & \multicolumn{6}{c}{\textbf{Legal Requirements}} \\
\cmidrule(lr){2-4} \cmidrule(lr){5-10}
 & & & & \multicolumn{2}{c|}{\textbf{Data Minimization}} & \textbf{Non-discrimination} & \textbf{Personal Data Qualification} & \multicolumn{2}{c}{\textbf{AML Requirements}} \\
\cmidrule(lr){5-10}
 & \textbf{Accuracy} & \textbf{Precision} & \textbf{F1 Score} & \textbf{\% Data Used} & \textbf{K-Anonymity} & \textbf{CDD (Gender)} & \textbf{Likelihood re-identification} & \textbf{Explainability} & \textbf{Recall} \\
\midrule
Set 1 & 0.85 & 0.80 & 0.86 & 84\% & No & 0.10 & Low & Moderate & 0.94 \\
Set 2 & 0.82 & 0.85 & 0.88 & 70\% & No & 0.12 & Low & High & 0.92 \\
\midrule
Set 3 & 0.83 & 0.79 & 0.85 & 70\% & Yes & 0.11 & Very Low & Moderate & 0.93 \\
Set 4 & 0.83 & 0.76 & 0.82 & 68\% & Yes & 0.13 & Very Low & High & 0.90 \\
\midrule
Set 5 & 0.82 & 0.78 & 0.84 & 68\% & No & 0.10 & Low & Moderate & 0.92 \\
Set 6 & 0.81 & 0.77 & 0.83 & 62\% & No & 0.06 & Low & High & 0.89 \\
\midrule
Set 7 & 0.84 & 0.79 & 0.84 & 72\% & Yes & 0.03 & Very Low & Moderate & 0.89 \\
Set 8 & 0.79 & 0.76 & 0.81 & 65\% & Yes & 0.07 & Very Low & High & 0.86 \\
\bottomrule
\end{tabular}%
}
\end{table*}

\subsection{Stage 5: Model Selection and Legal Justification}
The interdisciplinary team selects the model for deployment based on its performance and legal alignment, informed by the trade-off analysis from Stage 4. The chosen model must at least achieve reasonable scores on all proxy metrics or heuristics for the legal requirements, while maintaining high predictive performance, even if it does not excel in any single metric.

Our framework supports robust legal justification for the selected model. Documenting the rationale behind operationalizations, evaluation metrics (or heuristics), and proportionality-based trade-off decisions enables a compelling argument that the ML model can be deployed in line with legal obligations. This documentation clarifies the organization’s priorities and risk tolerances and provides necessary evidence for internal stakeholders, auditors, and regulatory authorities.

In the case study, the random forest model from Set 3 is chosen. Despite not excelling in any individual aspect, its overall positive evaluations outweigh slightly lower scores in other areas. The model's recall performance (0.93) aligns with the AML risk coverage requirement, mitigating the risk of fines, reputational damages, and potential revocation of its operating license. Although its recall is slightly lower than Set 1 (0.94) and similar to Sets 2 and 5 (0.92), Set 3’s alignment with data minimization and reduced risk of being classified as "personal data" through k-anonymity justifies its selection.
Set 3 shows significant gender-based treatment disparity (0.11). However, risks of non-discrimination law violations are considered limited, as some level of disparity is expected \citep{WACHTER2021105567} and is likely, at least partially, due to unmeasured circumstances. However, close monitoring is recommended.
While random forest models are less explainable than logistic regression, they offer moderate explainability, which is considered as legally justifiable. The choice between these models depends on balancing explainability with other legal obligations. In this case, better alignment with GDPR provisions favored Set 3.
Set 3 has slightly lower precision (0.79) compared to Set 1 (0.80) and significantly lower than Set 2 (0.85), indicating a higher false positive rate, offset by improved data protection.

In conclusion, this approach facilitates robust legal argumentation for key development choices and model selection. Decisions on operationalization of legal obligations are directly connected to model outcomes, revealing trade-offs with other legal aspects at the ML model level. At the same time, this design process also allows to achieve business objectives, but with due consideration for legal obligations.

\section{Conclusions and Future Work}

In this work, we presented an interdisciplinary five-step framework for organizations to develop and evaluate ML models amidst legal uncertainty. In this framework, the different operationalizations of each (or most) legal obligations are combined into unique combinations of operationalization sets. By developing models based on these sets, the framework yields explicit mappings of trade-offs at the ML model level which is indispensable input for a holistic assessment of the model's impact on legal obligations.

From the perspective of organizations, our framework is an important tool to fuel informed and legally justifiable decisions during ML model development, while safeguarding predictive performance. Moreover, after deployment, organizations can better understand the legal risks they could be exposed to when deploying a certain ML model. Finally, the rigorous documentation of the legal and, to a lesser extent, technical rationale behind the design of the final model facilitates compliance reporting.

From the perspective of the regulatory supervisors, this framework enhances oversight and facilitates improved supervisory guidance. It allows supervisors to assess the due diligence of organizations in aligning ML models with legal obligations. Recurrent themes in legal arguments and trade-offs can highlight areas needing further regulatory guidance or rule-making.

From the perspective of academics, our paper underscores the importance of interdisciplinary research for ensuring ML models are aligned with legal obligations. Future empirical studies can use our framework to explore the legal trade-offs in various settings (\textit{e.g.}, in the financial \citep{hoang2023machine, bockel2023causal}, healthcare \citep{giovanola2023beyond, gilbert2024eu,rasheed2022explainable} or governmental \citep{lewkowicz2024technical} sector), which can uncover insights into interactions between legal obligations at the ML model level in a given context.

\section*{Disclosure of generative AI usage}
In the preparation of this manuscript, the authors used ChatGPT and Claude AI in order to improve and refine language and to enhance readability. All AI-generated phrases were thoroughly reviewed and edited by the authors. The authors take full responsibility for the content of the publication.

\section*{Acknowledgments}
This project was supported by the FARI - AI for the Common Good Institute (ULB-VUB), financed by the European Union, with the support of the Brussels Capital Region (Innoviris and Paradigm).

\bibliographystyle{bib-style} 
\bibliography{EngineeringLawMLProblem}

\begin{thebibliography}{121}
\expandafter\ifx\csname natexlab\endcsname\relax\def\natexlab#1{#1}\fi
\providecommand{\url}[1]{\texttt{#1}}
\providecommand{\href}[2]{#2}
\providecommand{\path}[1]{#1}
\providecommand{\DOIprefix}{doi:}
\providecommand{\ArXivprefix}{arXiv:}
\providecommand{\URLprefix}{URL: }
\providecommand{\Pubmedprefix}{pmid:}
\providecommand{\doi}[1]{\href{http://dx.doi.org/#1}{\path{#1}}}
\providecommand{\Pubmed}[1]{\href{pmid:#1}{\path{#1}}}
\providecommand{\bibinfo}[2]{#2}
\ifx\xfnm\relax \def\xfnm[#1]{\unskip,\space#1}\fi
\bibitem[{Battaglini et~al.(2024)Battaglini, Guiso, Lacava, Miller, and Patacchini}]{battaglini2024refining}
\bibinfo{author}{M.~Battaglini}, \bibinfo{author}{L.~Guiso}, \bibinfo{author}{C.~Lacava}, \bibinfo{author}{D.~L. Miller}, \bibinfo{author}{E.~Patacchini},
\newblock \bibinfo{title}{Refining public policies with machine learning: The case of tax auditing},
\newblock \bibinfo{journal}{Journal of Econometrics}  (\bibinfo{year}{2024}) \bibinfo{pages}{105847}.
\bibitem[{De~Roux et~al.(2018)De~Roux, Perez, Moreno, Villamil, and Figueroa}]{de2018tax}
\bibinfo{author}{D.~De~Roux}, \bibinfo{author}{B.~Perez}, \bibinfo{author}{A.~Moreno}, \bibinfo{author}{M.~d.~P. Villamil}, \bibinfo{author}{C.~Figueroa},
\newblock \bibinfo{title}{Tax fraud detection for under-reporting declarations using an unsupervised machine learning approach},
\newblock in: \bibinfo{booktitle}{Proceedings of the 24th ACM SIGKDD International Conference on Knowledge Discovery \& Data Mining}, \bibinfo{year}{2018}, pp. \bibinfo{pages}{215--222}.
\bibitem[{Chen et~al.(2018)Chen, Van~Khoa, Teoh, Nazir, Karuppiah, and Lam}]{chen2018machine}
\bibinfo{author}{Z.~Chen}, \bibinfo{author}{L.~D. Van~Khoa}, \bibinfo{author}{E.~N. Teoh}, \bibinfo{author}{A.~Nazir}, \bibinfo{author}{E.~K. Karuppiah}, \bibinfo{author}{K.~S. Lam},
\newblock \bibinfo{title}{Machine learning techniques for anti-money laundering (aml) solutions in suspicious transaction detection: a review},
\newblock \bibinfo{journal}{Knowledge and Information Systems} \bibinfo{volume}{57} (\bibinfo{year}{2018}) \bibinfo{pages}{245--285}.
\bibitem[{Rahman and Kumar(2020)}]{rahman2020machine}
\bibinfo{author}{M.~Rahman}, \bibinfo{author}{V.~Kumar},
\newblock \bibinfo{title}{Machine learning based customer churn prediction in banking},
\newblock in: \bibinfo{booktitle}{2020 4th international conference on electronics, communication and aerospace technology (ICECA)}, \bibinfo{organization}{IEEE}, \bibinfo{year}{2020}, pp. \bibinfo{pages}{1196--1201}.
\bibitem[{De~Caigny et~al.(2024)De~Caigny, De~Bock, and Verboven}]{de2024hybrid}
\bibinfo{author}{A.~De~Caigny}, \bibinfo{author}{K.~W. De~Bock}, \bibinfo{author}{S.~Verboven},
\newblock \bibinfo{title}{Hybrid black-box classification for customer churn prediction with segmented interpretability analysis},
\newblock \bibinfo{journal}{Decision Support Systems} \bibinfo{volume}{181} (\bibinfo{year}{2024}) \bibinfo{pages}{114217}.
\bibitem[{Verboven and Martin(2021)}]{verboven2021combining}
\bibinfo{author}{S.~Verboven}, \bibinfo{author}{N.~Martin},
\newblock \bibinfo{title}{Combining the clinical and operational perspectives in heterogeneous treatment effect inference in healthcare processes},
\newblock in: \bibinfo{booktitle}{International Conference on Process Mining}, \bibinfo{organization}{Springer}, \bibinfo{year}{2021}, pp. \bibinfo{pages}{327--339}.
\bibitem[{Malgieri and Pasquale(2024)}]{malgieri2024licensing}
\bibinfo{author}{G.~Malgieri}, \bibinfo{author}{F.~Pasquale},
\newblock \bibinfo{title}{Licensing high-risk artificial intelligence: Toward ex ante justification for a disruptive technology},
\newblock \bibinfo{journal}{Computer Law \& Security Review} \bibinfo{volume}{52} (\bibinfo{year}{2024}) \bibinfo{pages}{105899}.
\bibitem[{Asgeirsson(2020)}]{asgeirsson2020nature}
\bibinfo{author}{H.~Asgeirsson}, \bibinfo{title}{The nature and value of vagueness in the law}, \bibinfo{publisher}{Bloomsbury Publishing}, \bibinfo{year}{2020}.
\bibitem[{Hildebrandt(2021)}]{hildebrandt2021adaptive}
\bibinfo{author}{M.~Hildebrandt},
\newblock \bibinfo{title}{The adaptive nature of text-driven law},
\newblock \bibinfo{journal}{Journal of Cross-disciplinary Research in Computational Law} \bibinfo{volume}{1} (\bibinfo{year}{2021}).
\bibitem[{Weerts et~al.(2023)Weerts, Xenidis, Tarissan, Olsen, and Pechenizkiy}]{weerts2023algorithmic}
\bibinfo{author}{H.~Weerts}, \bibinfo{author}{R.~Xenidis}, \bibinfo{author}{F.~Tarissan}, \bibinfo{author}{H.~P. Olsen}, \bibinfo{author}{M.~Pechenizkiy},
\newblock \bibinfo{title}{Algorithmic unfairness through the lens of eu non-discrimination law: Or why the law is not a decision tree},
\newblock in: \bibinfo{booktitle}{Proceedings of the 2023 ACM Conference on Fairness, Accountability, and Transparency}, \bibinfo{year}{2023}, pp. \bibinfo{pages}{805--816}.
\bibitem[{Dari-Mattiacci and Deffains(2007)}]{dari2007uncertainty}
\bibinfo{author}{G.~Dari-Mattiacci}, \bibinfo{author}{B.~Deffains},
\newblock \bibinfo{title}{Uncertainty of law and the legal process},
\newblock \bibinfo{journal}{Journal of Institutional and Theoretical Economics (JITE)/Zeitschrift f{\"u}r die gesamte Staatswissenschaft}  (\bibinfo{year}{2007}) \bibinfo{pages}{627--656}.
\bibitem[{Siena et~al.(2009)Siena, Mylopoulos, Perini, and Susi}]{siena2009designing}
\bibinfo{author}{A.~Siena}, \bibinfo{author}{J.~Mylopoulos}, \bibinfo{author}{A.~Perini}, \bibinfo{author}{A.~Susi},
\newblock \bibinfo{title}{Designing law-compliant software requirements},
\newblock in: \bibinfo{booktitle}{Conceptual Modeling-ER 2009: 28th International Conference on Conceptual Modeling, Gramado, Brazil, November 9-12, 2009. Proceedings 28}, \bibinfo{organization}{Springer}, \bibinfo{year}{2009}, pp. \bibinfo{pages}{472--486}.
\bibitem[{Otto and Ant{\'o}n(2007)}]{otto2007addressing}
\bibinfo{author}{P.~N. Otto}, \bibinfo{author}{A.~I. Ant{\'o}n},
\newblock \bibinfo{title}{Addressing legal requirements in requirements engineering},
\newblock in: \bibinfo{booktitle}{15th IEEE international requirements engineering conference (RE 2007)}, \bibinfo{organization}{IEEE}, \bibinfo{year}{2007}, pp. \bibinfo{pages}{5--14}.
\bibitem[{Calfee and Craswell(1984)}]{calfee1984some}
\bibinfo{author}{J.~E. Calfee}, \bibinfo{author}{R.~Craswell},
\newblock \bibinfo{title}{Some effects of uncertainty on compliance with legal standards},
\newblock \bibinfo{journal}{Va. L. Rev.} \bibinfo{volume}{70} (\bibinfo{year}{1984}) \bibinfo{pages}{965}.
\bibitem[{Hardt et~al.(2016)Hardt, Price, and Srebro}]{hardt2016equality}
\bibinfo{author}{M.~Hardt}, \bibinfo{author}{E.~Price}, \bibinfo{author}{N.~Srebro},
\newblock \bibinfo{title}{Equality of opportunity in supervised learning},
\newblock \bibinfo{journal}{Advances in neural information processing systems} \bibinfo{volume}{29} (\bibinfo{year}{2016}).
\bibitem[{Shanmugam et~al.(2022)Shanmugam, Diaz, Shabanian, Finck, and Biega}]{shanmugam2022learning}
\bibinfo{author}{D.~Shanmugam}, \bibinfo{author}{F.~Diaz}, \bibinfo{author}{S.~Shabanian}, \bibinfo{author}{M.~Finck}, \bibinfo{author}{A.~Biega},
\newblock \bibinfo{title}{Learning to limit data collection via scaling laws: A computational interpretation for the legal principle of data minimization},
\newblock in: \bibinfo{booktitle}{Proceedings of the 2022 ACM Conference on Fairness, Accountability, and Transparency}, \bibinfo{year}{2022}, pp. \bibinfo{pages}{839--849}.
\bibitem[{Goldsteen et~al.(2022)Goldsteen, Ezov, Shmelkin, Moffie, and Farkash}]{goldsteen2022data}
\bibinfo{author}{A.~Goldsteen}, \bibinfo{author}{G.~Ezov}, \bibinfo{author}{R.~Shmelkin}, \bibinfo{author}{M.~Moffie}, \bibinfo{author}{A.~Farkash},
\newblock \bibinfo{title}{Data minimization for gdpr compliance in machine learning models},
\newblock \bibinfo{journal}{AI and Ethics} \bibinfo{volume}{2} (\bibinfo{year}{2022}) \bibinfo{pages}{477--491}.
\bibitem[{Goldsteen et~al.(2023)Goldsteen, Saadi, Shmelkin, Shachor, and Razinkov}]{goldsteen2023ai}
\bibinfo{author}{A.~Goldsteen}, \bibinfo{author}{O.~Saadi}, \bibinfo{author}{R.~Shmelkin}, \bibinfo{author}{S.~Shachor}, \bibinfo{author}{N.~Razinkov},
\newblock \bibinfo{title}{Ai privacy toolkit},
\newblock \bibinfo{journal}{SoftwareX} \bibinfo{volume}{22} (\bibinfo{year}{2023}) \bibinfo{pages}{101352}.
\bibitem[{Rudin(2019)}]{rudin2019stop}
\bibinfo{author}{C.~Rudin},
\newblock \bibinfo{title}{Stop explaining black box machine learning models for high stakes decisions and use interpretable models instead},
\newblock \bibinfo{journal}{Nature machine intelligence} \bibinfo{volume}{1} (\bibinfo{year}{2019}) \bibinfo{pages}{206--215}.
\bibitem[{Fresz et~al.(2024)Fresz, Dubovitskaya, Brajovic, Huber, and Horz}]{fresz2024should}
\bibinfo{author}{B.~Fresz}, \bibinfo{author}{E.~Dubovitskaya}, \bibinfo{author}{D.~Brajovic}, \bibinfo{author}{M.~F. Huber}, \bibinfo{author}{C.~Horz},
\newblock \bibinfo{title}{How should ai decisions be explained? requirements for explanations from the perspective of european law},
\newblock in: \bibinfo{booktitle}{Proceedings of the AAAI/ACM Conference on AI, Ethics, and Society}, volume~\bibinfo{volume}{7}, \bibinfo{year}{2024}, pp. \bibinfo{pages}{438--450}.
\bibitem[{Wachter et~al.(2021)Wachter, Mittelstadt, and Russell}]{WACHTER2021105567}
\bibinfo{author}{S.~Wachter}, \bibinfo{author}{B.~Mittelstadt}, \bibinfo{author}{C.~Russell},
\newblock \bibinfo{title}{Why fairness cannot be automated: Bridging the gap between eu non-discrimination law and ai},
\newblock \bibinfo{journal}{Computer Law \& Security Review} \bibinfo{volume}{41} (\bibinfo{year}{2021}) \bibinfo{pages}{105567}. \URLprefix \url{https://www.sciencedirect.com/science/article/pii/S0267364921000406}. \DOIprefix\doi{https://doi.org/10.1016/j.clsr.2021.105567}.
\bibitem[{Bandara et~al.(2020)Bandara, Bandara, and Fernando}]{bandara2020evaluation}
\bibinfo{author}{P.~K. Bandara}, \bibinfo{author}{H.~D. Bandara}, \bibinfo{author}{S.~Fernando},
\newblock \bibinfo{title}{Evaluation of re-identification risks in data anonymization techniques based on population uniqueness},
\newblock in: \bibinfo{booktitle}{2020 5th International Conference on Information Technology Research (ICITR)}, \bibinfo{organization}{IEEE}, \bibinfo{year}{2020}, pp. \bibinfo{pages}{1--5}.
\bibitem[{Agarwal(2021)}]{agarwal2021tradepriv}
\bibinfo{author}{S.~Agarwal},
\newblock \bibinfo{title}{Trade-offs between fairness and privacy in machine learning},
\newblock in: \bibinfo{booktitle}{IJCAI 2021 Workshop on AI for Social Good}, \bibinfo{year}{2021}.
\bibitem[{Gu et~al.(2022)Gu, Tianqing, Li, Zhang, Ren, and Choo}]{gu2022privacy}
\bibinfo{author}{X.~Gu}, \bibinfo{author}{Z.~Tianqing}, \bibinfo{author}{J.~Li}, \bibinfo{author}{T.~Zhang}, \bibinfo{author}{W.~Ren}, \bibinfo{author}{K.-K.~R. Choo},
\newblock \bibinfo{title}{Privacy, accuracy, and model fairness trade-offs in federated learning},
\newblock \bibinfo{journal}{Computers \& Security} \bibinfo{volume}{122} (\bibinfo{year}{2022}) \bibinfo{pages}{102907}.
\bibitem[{Sun et~al.(2023)Sun, Zhang, Lin, Li, Wang, and Li}]{sun2023tradeoffs}
\bibinfo{author}{K.~Sun}, \bibinfo{author}{X.~Zhang}, \bibinfo{author}{X.~Lin}, \bibinfo{author}{G.~Li}, \bibinfo{author}{J.~Wang}, \bibinfo{author}{J.~Li}, \bibinfo{title}{Toward the tradeoffs between privacy, fairness and utility in federated learning}, \bibinfo{year}{2023}. \href{http://arxiv.org/abs/2311.18190}{{\tt arXiv:2311.18190}}.
\bibitem[{Ferry et~al.(2023)Ferry, A{\"\i}vodji, Gambs, Huguet, and Siala}]{ferry2023sok}
\bibinfo{author}{J.~Ferry}, \bibinfo{author}{U.~A{\"\i}vodji}, \bibinfo{author}{S.~Gambs}, \bibinfo{author}{M.-J. Huguet}, \bibinfo{author}{M.~Siala},
\newblock \bibinfo{title}{Sok: Taming the triangle--on the interplays between fairness, interpretability and privacy in machine learning},
\newblock \bibinfo{journal}{arXiv preprint arXiv:2312.16191}  (\bibinfo{year}{2023}).
\bibitem[{Pawelczyk et~al.(2023)Pawelczyk, Leemann, Biega, and Kasneci}]{pawelczyk2023tradeoff}
\bibinfo{author}{M.~Pawelczyk}, \bibinfo{author}{T.~Leemann}, \bibinfo{author}{A.~Biega}, \bibinfo{author}{G.~Kasneci}, \bibinfo{title}{On the trade-off between actionable explanations and the right to be forgotten}, \bibinfo{year}{2023}. \href{http://arxiv.org/abs/2208.14137}{{\tt arXiv:2208.14137}}.
\bibitem[{Xhemajli(2021)}]{xhemajli2021role}
\bibinfo{author}{H.~Xhemajli},
\newblock \bibinfo{title}{The role of ethics and morality in law: Similarities and differences},
\newblock \bibinfo{journal}{Ohio NUL Rev.} \bibinfo{volume}{48} (\bibinfo{year}{2021}) \bibinfo{pages}{81}.
\bibitem[{Gardner(2010)}]{gardner201035}
\bibinfo{author}{J.~Gardner},
\newblock \bibinfo{title}{35 ethics and law},
\newblock \bibinfo{journal}{The Routledge Companion to Ethics}  (\bibinfo{year}{2010}).
\bibitem[{Rentmeester(2006)}]{rentmeester2006s}
\bibinfo{author}{C.~A. Rentmeester},
\newblock \bibinfo{title}{What's legal? what's moral? what's the difference? a guide for teaching residents},
\newblock \bibinfo{journal}{The American Journal of Bioethics} \bibinfo{volume}{6} (\bibinfo{year}{2006}) \bibinfo{pages}{31--33}.
\bibitem[{Lewkowicz and Sarf(2024)}]{lewkowicz2024technical}
\bibinfo{author}{G.~Lewkowicz}, \bibinfo{author}{R.~Sarf},
\newblock \bibinfo{title}{Taking technical standardization of fundamental rights seriously for trustworthy artificial intelligence},
\newblock \bibinfo{journal}{La Revue des Juristes de Sciences Po}  (\bibinfo{year}{2024}) \bibinfo{pages}{42--46}. \URLprefix \url{https://centreperelman.be//content/uploads/2024/03/GL-RS-AI-Technical-standards-and-fundamental-rights.pdf}.
\bibitem[{Selbst et~al.(2019)Selbst, Boyd, Friedler, Venkatasubramanian, and Vertesi}]{selbst2019fairness}
\bibinfo{author}{A.~D. Selbst}, \bibinfo{author}{D.~Boyd}, \bibinfo{author}{S.~A. Friedler}, \bibinfo{author}{S.~Venkatasubramanian}, \bibinfo{author}{J.~Vertesi},
\newblock \bibinfo{title}{Fairness and abstraction in sociotechnical systems},
\newblock in: \bibinfo{booktitle}{Proceedings of the conference on fairness, accountability, and transparency}, \bibinfo{year}{2019}, pp. \bibinfo{pages}{59--68}.
\bibitem[{Blackman(2022)}]{blackman2022ethical}
\bibinfo{author}{R.~Blackman}, \bibinfo{title}{Ethical machines: Your concise guide to totally unbiased, transparent, and respectful AI}, \bibinfo{publisher}{Harvard Business Press}, \bibinfo{year}{2022}.
\bibitem[{Koumeri et~al.(2023)Koumeri, Legast, Yousefi, Vanhoof, Legay, and Schommer}]{koumeri2023compatibility}
\bibinfo{author}{L.~K. Koumeri}, \bibinfo{author}{M.~Legast}, \bibinfo{author}{Y.~Yousefi}, \bibinfo{author}{K.~Vanhoof}, \bibinfo{author}{A.~Legay}, \bibinfo{author}{C.~Schommer},
\newblock \bibinfo{title}{Compatibility of fairness metrics with eu non-discrimination laws: Demographic parity \& conditional demographic disparity},
\newblock \bibinfo{journal}{arXiv preprint arXiv:2306.08394}  (\bibinfo{year}{2023}).
\bibitem[{Kerrigan(2022)}]{kerrigan2022artificial}
\bibinfo{author}{C.~Kerrigan}, \bibinfo{title}{Artificial intelligence: Law and regulation}, \bibinfo{publisher}{Edward Elgar Publishing}, \bibinfo{year}{2022}.
\bibitem[{Hadjiemmanuil(2015)}]{heavily-regulated-industry}
\bibinfo{author}{C.~Hadjiemmanuil},
\newblock \bibinfo{title}{A heavily regulated industry},
\newblock \bibinfo{journal}{eucrim}  (\bibinfo{year}{2015}).
\bibitem[{Shen(2023)}]{shen2023ai}
\bibinfo{author}{N.~Shen},
\newblock \bibinfo{title}{Ai regulation in health care: How washington state can conquer the new territory of ai regulation},
\newblock \bibinfo{journal}{Seattle J. Tech. Env't \& Innovation L.} \bibinfo{volume}{13} (\bibinfo{year}{2023}) \bibinfo{pages}{1}.
\bibitem[{Hassani et~al.(2024)Hassani, Sabetzadeh, Amyot, and Liao}]{hassani2024rethinking}
\bibinfo{author}{S.~Hassani}, \bibinfo{author}{M.~Sabetzadeh}, \bibinfo{author}{D.~Amyot}, \bibinfo{author}{J.~Liao},
\newblock \bibinfo{title}{Rethinking legal compliance automation: Opportunities with large language models},
\newblock \bibinfo{journal}{arXiv preprint arXiv:2404.14356}  (\bibinfo{year}{2024}).
\bibitem[{Sleimi et~al.(2019)Sleimi, Ceci, Sannier, Sabetzadeh, Briand, and Dann}]{sleimi2019query}
\bibinfo{author}{A.~Sleimi}, \bibinfo{author}{M.~Ceci}, \bibinfo{author}{N.~Sannier}, \bibinfo{author}{M.~Sabetzadeh}, \bibinfo{author}{L.~Briand}, \bibinfo{author}{J.~Dann},
\newblock \bibinfo{title}{A query system for extracting requirements-related information from legal texts},
\newblock in: \bibinfo{booktitle}{2019 IEEE 27th international requirements engineering conference (RE)}, \bibinfo{organization}{IEEE}, \bibinfo{year}{2019}, pp. \bibinfo{pages}{319--329}.
\bibitem[{Akhigbe et~al.(2019)Akhigbe, Amyot, and Richards}]{akhigbe2019systematic}
\bibinfo{author}{O.~Akhigbe}, \bibinfo{author}{D.~Amyot}, \bibinfo{author}{G.~Richards},
\newblock \bibinfo{title}{A systematic literature mapping of goal and non-goal modelling methods for legal and regulatory compliance},
\newblock \bibinfo{journal}{Requirements Engineering} \bibinfo{volume}{24} (\bibinfo{year}{2019}) \bibinfo{pages}{459--481}.
\bibitem[{Abualhaija et~al.(2023)Abualhaija, Ceci, and Briand}]{abualhaija2023legal}
\bibinfo{author}{S.~Abualhaija}, \bibinfo{author}{M.~Ceci}, \bibinfo{author}{L.~Briand},
\newblock \bibinfo{title}{Legal requirements analysis},
\newblock \bibinfo{journal}{arXiv preprint arXiv:2311.13871}  (\bibinfo{year}{2023}).
\bibitem[{Fa{\ss}bender(2017)}]{fassbender2017domain}
\bibinfo{author}{S.~Fa{\ss}bender}, \bibinfo{title}{Domain-and Quality-aware Requirements Engineering for Law-compliant Systems}, Ph.D. thesis, Dissertation, Duisburg, Essen, Universit{\"a}t Duisburg-Essen, 2017, \bibinfo{year}{2017}.
\bibitem[{Hoess et~al.(2024)Hoess, Pocher, Roth, and Fridgen}]{hoess2024towards}
\bibinfo{author}{A.~Hoess}, \bibinfo{author}{N.~Pocher}, \bibinfo{author}{T.~H. Roth}, \bibinfo{author}{G.~Fridgen},
\newblock \bibinfo{title}{Towards a design science research process for legal compliance by design},
\newblock in: \bibinfo{booktitle}{PACIS 2024 Proceedings}, \bibinfo{year}{2024}, p. \bibinfo{pages}{1609}.
\bibitem[{Gjorgjevikj et~al.(2023)Gjorgjevikj, Mishev, Antovski, and Trajanov}]{gjorgjevikj2023requirements}
\bibinfo{author}{A.~Gjorgjevikj}, \bibinfo{author}{K.~Mishev}, \bibinfo{author}{L.~Antovski}, \bibinfo{author}{D.~Trajanov},
\newblock \bibinfo{title}{Requirements engineering in machine learning projects},
\newblock \bibinfo{journal}{IEEE Access}  (\bibinfo{year}{2023}).
\bibitem[{Zowghi and Bano(2023)}]{zowghi2023s}
\bibinfo{author}{D.~Zowghi}, \bibinfo{author}{M.~Bano},
\newblock \bibinfo{title}{What’s missing in requirements engineering for responsible ai?},
\newblock \bibinfo{journal}{IEEE Software} \bibinfo{volume}{40} (\bibinfo{year}{2023}) \bibinfo{pages}{11--15}.
\bibitem[{Budiardjo et~al.(2024)}]{budiardjo2024roadmap}
\bibinfo{author}{E.~K. Budiardjo}, et~al.,
\newblock \bibinfo{title}{Roadmap analysis of artificial intelligence engineering method.},
\newblock \bibinfo{journal}{Revue d'Intelligence Artificielle} \bibinfo{volume}{38} (\bibinfo{year}{2024}).
\bibitem[{Ahmad et~al.(2023)Ahmad, Abdelrazek, Arora, Bano, and Grundy}]{ahmad2023requirements}
\bibinfo{author}{K.~Ahmad}, \bibinfo{author}{M.~Abdelrazek}, \bibinfo{author}{C.~Arora}, \bibinfo{author}{M.~Bano}, \bibinfo{author}{J.~Grundy},
\newblock \bibinfo{title}{Requirements engineering for artificial intelligence systems: A systematic mapping study},
\newblock \bibinfo{journal}{Information and Software Technology} \bibinfo{volume}{158} (\bibinfo{year}{2023}) \bibinfo{pages}{107176}.
\bibitem[{Mart{\'\i}nez-Fern{\'a}ndez et~al.(2022)Mart{\'\i}nez-Fern{\'a}ndez, Bogner, Franch, Oriol, Siebert, Trendowicz, Vollmer, and Wagner}]{martinez2022software}
\bibinfo{author}{S.~Mart{\'\i}nez-Fern{\'a}ndez}, \bibinfo{author}{J.~Bogner}, \bibinfo{author}{X.~Franch}, \bibinfo{author}{M.~Oriol}, \bibinfo{author}{J.~Siebert}, \bibinfo{author}{A.~Trendowicz}, \bibinfo{author}{A.~M. Vollmer}, \bibinfo{author}{S.~Wagner},
\newblock \bibinfo{title}{Software engineering for ai-based systems: a survey},
\newblock \bibinfo{journal}{ACM Transactions on Software Engineering and Methodology (TOSEM)} \bibinfo{volume}{31} (\bibinfo{year}{2022}) \bibinfo{pages}{1--59}.
\bibitem[{De~Martino and Palomba(2023)}]{de2023classification}
\bibinfo{author}{V.~De~Martino}, \bibinfo{author}{F.~Palomba},
\newblock \bibinfo{title}{Classification, challenges, and automated approaches to handle non-functional requirements in ml-enabled systems: A systematic literature review},
\newblock \bibinfo{journal}{arXiv preprint arXiv:2311.17483}  (\bibinfo{year}{2023}).
\bibitem[{Boella et~al.(2014)Boella, Humphreys, Muthuri, Rossi, and van~der Torre}]{boella2014critical}
\bibinfo{author}{G.~Boella}, \bibinfo{author}{L.~Humphreys}, \bibinfo{author}{R.~Muthuri}, \bibinfo{author}{P.~Rossi}, \bibinfo{author}{L.~van~der Torre},
\newblock \bibinfo{title}{A critical analysis of legal requirements engineering from the perspective of legal practice},
\newblock in: \bibinfo{booktitle}{2014 IEEE 7th International Workshop on Requirements Engineering and Law (RELAW)}, \bibinfo{organization}{IEEE}, \bibinfo{year}{2014}, pp. \bibinfo{pages}{14--21}.
\bibitem[{IEEE(1990)}]{159342}
\bibinfo{author}{IEEE}, \bibinfo{title}{Ieee standard glossary of software engineering terminology}, \bibinfo{year}{1990}. \DOIprefix\doi{10.1109/IEEESTD.1990.101064}.
\bibitem[{Van~Vliet et~al.(2008)Van~Vliet, Van~Vliet, and Van~Vliet}]{van2008software}
\bibinfo{author}{H.~Van~Vliet}, \bibinfo{author}{H.~Van~Vliet}, \bibinfo{author}{J.~Van~Vliet}, \bibinfo{title}{Software engineering: principles and practice}, volume~\bibinfo{volume}{13}, \bibinfo{publisher}{John Wiley \& Sons Hoboken, NJ}, \bibinfo{year}{2008}.
\bibitem[{Belani et~al.(2019)Belani, Vukovic, and Car}]{belani2019requirements}
\bibinfo{author}{H.~Belani}, \bibinfo{author}{M.~Vukovic}, \bibinfo{author}{{\v{Z}}.~Car},
\newblock \bibinfo{title}{Requirements engineering challenges in building ai-based complex systems},
\newblock in: \bibinfo{booktitle}{2019 IEEE 27th International Requirements Engineering Conference Workshops (REW)}, \bibinfo{organization}{IEEE}, \bibinfo{year}{2019}, pp. \bibinfo{pages}{252--255}.
\bibitem[{Hildebrandt and De~Bois(2021)}]{hildebrandt2021law}
\bibinfo{author}{M.~Hildebrandt}, \bibinfo{author}{A.~De~Bois},
\newblock \bibinfo{title}{Law for computer scientists},
\newblock in: \bibinfo{booktitle}{ECCAI Advanced Course on Artificial Intelligence}, \bibinfo{publisher}{Springer}, \bibinfo{year}{2021}, pp. \bibinfo{pages}{261--274}.
\bibitem[{Merriam-Webster(2025)}]{mw:law}
\bibinfo{author}{Merriam-Webster}, \bibinfo{title}{Merriam-webster.com dictionary}, \bibinfo{howpublished}{\url{https://www.merriam-webster.com/dictionary/law}}, \bibinfo{year}{2025}.
\bibitem[{Ghanavati et~al.(2016)Ghanavati, Amyot, Siena, Susi, and Perini}]{ghanavati2016making}
\bibinfo{author}{S.~Ghanavati}, \bibinfo{author}{D.~Amyot}, \bibinfo{author}{A.~Siena}, \bibinfo{author}{A.~Susi}, \bibinfo{author}{A.~Perini},
\newblock \bibinfo{title}{Making business processes law compliant},
\newblock in: \bibinfo{booktitle}{First workshop on law compliancy issues in organisational systems and strategies (iComply’10). Retrieved}, volume~\bibinfo{volume}{5}, \bibinfo{year}{2016}.
\bibitem[{Castellanos~Ardila et~al.(2022)Castellanos~Ardila, Gallina, and Ul~Muram}]{castellanos2022compliance}
\bibinfo{author}{J.~P. Castellanos~Ardila}, \bibinfo{author}{B.~Gallina}, \bibinfo{author}{F.~Ul~Muram},
\newblock \bibinfo{title}{Compliance checking of software processes: A systematic literature review},
\newblock \bibinfo{journal}{Journal of Software: Evolution and Process} \bibinfo{volume}{34} (\bibinfo{year}{2022}) \bibinfo{pages}{e2440}.
\bibitem[{Ahmad et~al.(2023{\natexlab{a}})Ahmad, Abdelrazek, Arora, Grundy, and Bano}]{ahmad2023requirementsEmpirical}
\bibinfo{author}{K.~Ahmad}, \bibinfo{author}{M.~Abdelrazek}, \bibinfo{author}{C.~Arora}, \bibinfo{author}{J.~Grundy}, \bibinfo{author}{M.~Bano},
\newblock \bibinfo{title}{Requirements elicitation and modelling of artificial intelligence systems: An empirical study},
\newblock \bibinfo{journal}{arXiv preprint arXiv:2302.06034}  (\bibinfo{year}{2023}{\natexlab{a}}).
\bibitem[{Ahmad et~al.(2023{\natexlab{b}})Ahmad, Abdelrazek, Arora, Baniya, Bano, and Grundy}]{ahmad2023requirementsFramework}
\bibinfo{author}{K.~Ahmad}, \bibinfo{author}{M.~Abdelrazek}, \bibinfo{author}{C.~Arora}, \bibinfo{author}{A.~A. Baniya}, \bibinfo{author}{M.~Bano}, \bibinfo{author}{J.~Grundy},
\newblock \bibinfo{title}{Requirements engineering framework for human-centered artificial intelligence software systems},
\newblock \bibinfo{journal}{Applied Soft Computing} \bibinfo{volume}{143} (\bibinfo{year}{2023}{\natexlab{b}}) \bibinfo{pages}{110455}.
\bibitem[{Ahmad et~al.(2023{\natexlab{c}})Ahmad, Arora, Abdelrazek, Grundy, and Vasa}]{ahmad2023requirementsElicitation}
\bibinfo{author}{K.~Ahmad}, \bibinfo{author}{C.~Arora}, \bibinfo{author}{M.~Abdelrazek}, \bibinfo{author}{J.~Grundy}, \bibinfo{author}{R.~Vasa},
\newblock \bibinfo{title}{Requirements elicitation in the age of ai: A tool’s multi-system journey},
\newblock in: \bibinfo{booktitle}{International Conference on Evaluation of Novel Approaches to Software Engineering}, \bibinfo{organization}{Springer}, \bibinfo{year}{2023}{\natexlab{c}}, pp. \bibinfo{pages}{67--90}.
\bibitem[{Bano et~al.(2023)Bano, Zowghi, Gervasi, and Shams}]{bano2023ai}
\bibinfo{author}{M.~Bano}, \bibinfo{author}{D.~Zowghi}, \bibinfo{author}{V.~Gervasi}, \bibinfo{author}{R.~Shams},
\newblock \bibinfo{title}{Ai for all: Operationalising diversity and inclusion requirements for ai systems},
\newblock \bibinfo{journal}{arXiv preprint arXiv:2311.14695}  (\bibinfo{year}{2023}).
\bibitem[{Sporsem(2024)}]{sporsem2024discovering}
\bibinfo{author}{T.~Sporsem},
\newblock \bibinfo{title}{Discovering explainability requirements in ml-based software},
\newblock in: \bibinfo{booktitle}{Proceedings of the 2024 IEEE/ACM 46th International Conference on Software Engineering: Companion Proceedings}, \bibinfo{year}{2024}, pp. \bibinfo{pages}{170--172}.
\bibitem[{Balasubramaniam et~al.(2024)Balasubramaniam, Kauppinen, Truong, and Kujala}]{balasubramaniam2024candidate}
\bibinfo{author}{N.~Balasubramaniam}, \bibinfo{author}{M.~Kauppinen}, \bibinfo{author}{H.-L. Truong}, \bibinfo{author}{S.~Kujala},
\newblock \bibinfo{title}{Candidate solutions for defining explainability requirements of ai systems},
\newblock in: \bibinfo{booktitle}{International Working Conference on Requirements Engineering: Foundation for Software Quality}, \bibinfo{organization}{Springer}, \bibinfo{year}{2024}, pp. \bibinfo{pages}{129--146}.
\bibitem[{Damirchi and Amini(2023)}]{damirchi2023non}
\bibinfo{author}{R.~Damirchi}, \bibinfo{author}{A.~Amini},
\newblock \bibinfo{title}{Non-functional requirement extracting methods for ai-based systems: A survey},
\newblock in: \bibinfo{booktitle}{2023 13th International Conference on Computer and Knowledge Engineering (ICCKE)}, \bibinfo{organization}{IEEE}, \bibinfo{year}{2023}, pp. \bibinfo{pages}{535--539}.
\bibitem[{Martins and Novo(2024)}]{martinsrequirements}
\bibinfo{author}{M.~C. Martins}, \bibinfo{author}{T.~Novo},
\newblock \bibinfo{title}{A requirements engineering process for machine learning innovation projects},
\newblock in: \bibinfo{booktitle}{Proceedings of the 27th Workshop on Requirements Engineering (WER24)}, \bibinfo{year}{2024}.
\bibitem[{Sadovski et~al.(2024)Sadovski, Aviv, and Hadar}]{sadovski2024towards}
\bibinfo{author}{E.~Sadovski}, \bibinfo{author}{I.~Aviv}, \bibinfo{author}{I.~Hadar},
\newblock \bibinfo{title}{Towards a comprehensive ontology for requirements engineering for ai-powered systems},
\newblock in: \bibinfo{booktitle}{International Working Conference on Requirements Engineering: Foundation for Software Quality}, \bibinfo{organization}{Springer}, \bibinfo{year}{2024}, pp. \bibinfo{pages}{219--230}.
\bibitem[{Agarwal(2021)}]{agarwal2021tradeinterpret}
\bibinfo{author}{S.~Agarwal},
\newblock \bibinfo{title}{Trade-offs between fairness and interpretability in machine learning},
\newblock in: \bibinfo{booktitle}{IJCAI 2021 Workshop on AI for Social Good}, \bibinfo{year}{2021}, pp. \bibinfo{pages}{1--6}.
\bibitem[{Jabbari et~al.(2020)Jabbari, Ou, Lakkaraju, and Tambe}]{jabbari2020empirical}
\bibinfo{author}{S.~Jabbari}, \bibinfo{author}{H.-C. Ou}, \bibinfo{author}{H.~Lakkaraju}, \bibinfo{author}{M.~Tambe},
\newblock \bibinfo{title}{An empirical study of the trade-offs between interpretability and fairness},
\newblock in: \bibinfo{booktitle}{ICML Workshop on Human Interpretability in Machine Learning, International Conference on Machine Learning (ICML)}, \bibinfo{year}{2020}.
\bibitem[{Jo et~al.(2022)Jo, Aghaei, G{\'o}mez, and Vayanos}]{jo2022learning}
\bibinfo{author}{N.~Jo}, \bibinfo{author}{S.~Aghaei}, \bibinfo{author}{A.~G{\'o}mez}, \bibinfo{author}{P.~Vayanos},
\newblock \bibinfo{title}{Learning optimal fair classification trees: Trade-offs between interpretability, fairness, and accuracy},
\newblock \bibinfo{journal}{arXiv preprint arXiv:2201.09932}  (\bibinfo{year}{2022}).
\bibitem[{Pannekoek and Spigler(2021)}]{pannekoek2021investigating}
\bibinfo{author}{M.~Pannekoek}, \bibinfo{author}{G.~Spigler},
\newblock \bibinfo{title}{Investigating trade-offs in utility, fairness and differential privacy in neural networks},
\newblock \bibinfo{journal}{arXiv preprint arXiv:2102.05975}  (\bibinfo{year}{2021}).
\bibitem[{Gittens et~al.(2022)Gittens, Yener, and Yung}]{gittens2022adversarial}
\bibinfo{author}{A.~Gittens}, \bibinfo{author}{B.~Yener}, \bibinfo{author}{M.~Yung},
\newblock \bibinfo{title}{An adversarial perspective on accuracy, robustness, fairness, and privacy: Multilateral-tradeoffs in trustworthy ml},
\newblock \bibinfo{journal}{IEEE Access} \bibinfo{volume}{10} (\bibinfo{year}{2022}) \bibinfo{pages}{120850--120865}.
\bibitem[{Rodolfa et~al.(2021)Rodolfa, Lamba, and Ghani}]{rodolfa2021empirical}
\bibinfo{author}{K.~T. Rodolfa}, \bibinfo{author}{H.~Lamba}, \bibinfo{author}{R.~Ghani},
\newblock \bibinfo{title}{Empirical observation of negligible fairness--accuracy trade-offs in machine learning for public policy},
\newblock \bibinfo{journal}{Nature Machine Intelligence} \bibinfo{volume}{3} (\bibinfo{year}{2021}) \bibinfo{pages}{896--904}.
\bibitem[{Finck and Biega(2021)}]{https://doi.org/10.26116/techreg.2021.004}
\bibinfo{author}{M.~Finck}, \bibinfo{author}{A.~J. Biega},
\newblock \bibinfo{title}{Reviving purpose limitation and data minimisation in data-driven systems},
\newblock \bibinfo{journal}{Technology and Regulation}  (\bibinfo{year}{2021}) \bibinfo{pages}{Vol. 2021 (2021)}. \URLprefix \url{https://techreg.org/article/view/10986}. \DOIprefix\doi{10.26116/TECHREG.2021.004}.
\bibitem[{Pessach and Shmueli(2022)}]{pessach2022review}
\bibinfo{author}{D.~Pessach}, \bibinfo{author}{E.~Shmueli},
\newblock \bibinfo{title}{A review on fairness in machine learning},
\newblock \bibinfo{journal}{ACM Computing Surveys (CSUR)} \bibinfo{volume}{55} (\bibinfo{year}{2022}) \bibinfo{pages}{1--44}.
\bibitem[{Vandevelde(2018)}]{vandevelde2018thinking}
\bibinfo{author}{K.~J. Vandevelde}, \bibinfo{title}{Thinking like a lawyer: An introduction to legal reasoning}, \bibinfo{publisher}{Routledge}, \bibinfo{year}{2018}.
\bibitem[{Bueno~Mom{\v{c}}ilovi{\'c} and Balta(2024)}]{bueno2024challenges}
\bibinfo{author}{T.~Bueno~Mom{\v{c}}ilovi{\'c}}, \bibinfo{author}{D.~Balta},
\newblock \bibinfo{title}{Challenges of assuring compliance of information systems in finance},
\newblock in: \bibinfo{booktitle}{International Conference on Software Quality}, \bibinfo{organization}{Springer}, \bibinfo{year}{2024}, pp. \bibinfo{pages}{135--152}.
\bibitem[{Kaminski and Malgieri(2020)}]{kaminski2020algorithmic}
\bibinfo{author}{M.~E. Kaminski}, \bibinfo{author}{G.~Malgieri}, \bibinfo{title}{Algorithmic impact assessments under the GDPR: producing multi-layered explanations}, \bibinfo{publisher}{HeinOnline}, \bibinfo{year}{2020}.
\bibitem[{Goodman and Flaxman(2017)}]{goodman2017european}
\bibinfo{author}{B.~Goodman}, \bibinfo{author}{S.~Flaxman},
\newblock \bibinfo{title}{European union regulations on algorithmic decision-making and a “right to explanation”},
\newblock \bibinfo{journal}{AI magazine} \bibinfo{volume}{38} (\bibinfo{year}{2017}) \bibinfo{pages}{50--57}.
\bibitem[{Selbst and Powles(2018)}]{selbst2018meaningful}
\bibinfo{author}{A.~Selbst}, \bibinfo{author}{J.~Powles},
\newblock \bibinfo{title}{“meaningful information” and the right to explanation},
\newblock in: \bibinfo{booktitle}{conference on fairness, accountability and transparency}, \bibinfo{organization}{PMLR}, \bibinfo{year}{2018}, pp. \bibinfo{pages}{48--48}.
\bibitem[{Malgieri and Comand{\'e}(2017)}]{malgieri2017right}
\bibinfo{author}{G.~Malgieri}, \bibinfo{author}{G.~Comand{\'e}},
\newblock \bibinfo{title}{Why a right to legibility of automated decision-making exists in the general data protection regulation},
\newblock \bibinfo{journal}{International Data Privacy Law} \bibinfo{volume}{7} (\bibinfo{year}{2017}) \bibinfo{pages}{243--265}.
\bibitem[{Bayaml{\i}o{\u{g}}lu(2022)}]{bayamliouglu2022right}
\bibinfo{author}{E.~Bayaml{\i}o{\u{g}}lu},
\newblock \bibinfo{title}{The right to contest automated decisions under the general data protection regulation: Beyond the so-called “right to explanation”},
\newblock \bibinfo{journal}{Regulation \& Governance} \bibinfo{volume}{16} (\bibinfo{year}{2022}) \bibinfo{pages}{1058--1078}.
\bibitem[{Wachter et~al.(2017)Wachter, Mittelstadt, and Floridi}]{wachter2017right}
\bibinfo{author}{S.~Wachter}, \bibinfo{author}{B.~Mittelstadt}, \bibinfo{author}{L.~Floridi},
\newblock \bibinfo{title}{Why a right to explanation of automated decision-making does not exist in the general data protection regulation},
\newblock \bibinfo{journal}{International data privacy law} \bibinfo{volume}{7} (\bibinfo{year}{2017}) \bibinfo{pages}{76--99}.
\bibitem[{Custers(2022)}]{custers2022new}
\bibinfo{author}{B.~Custers},
\newblock \bibinfo{title}{New digital rights: Imagining additional fundamental rights for the digital era},
\newblock \bibinfo{journal}{Computer Law \& Security Review} \bibinfo{volume}{44} (\bibinfo{year}{2022}) \bibinfo{pages}{105636}.
\bibitem[{Tridimas(2023)}]{tridimas2023wreaking}
\bibinfo{author}{T.~Tridimas},
\newblock \bibinfo{title}{Wreaking the wrongs: Balancing rights and the public interest the eu way},
\newblock \bibinfo{journal}{Colum. J. Eur. L.} \bibinfo{volume}{29} (\bibinfo{year}{2023}) \bibinfo{pages}{185}.
\bibitem[{Gellert(2016)}]{gellert2016we}
\bibinfo{author}{R.~Gellert},
\newblock \bibinfo{title}{We have always managed risks in data protection law: understanding the similarities and differences between the rights-based and the risk-based approaches to data protection},
\newblock \bibinfo{journal}{Eur. Data Prot. L. Rev.} \bibinfo{volume}{2} (\bibinfo{year}{2016}) \bibinfo{pages}{481}.
\bibitem[{Lin(2019)}]{lin2019artificial}
\bibinfo{author}{T.~C. Lin},
\newblock \bibinfo{title}{Artificial intelligence, finance, and the law},
\newblock \bibinfo{journal}{Fordham L. Rev.} \bibinfo{volume}{88} (\bibinfo{year}{2019}) \bibinfo{pages}{531}.
\bibitem[{Jensen and Iosifidis(2023)}]{jensen2023fighting}
\bibinfo{author}{R.~I.~T. Jensen}, \bibinfo{author}{A.~Iosifidis},
\newblock \bibinfo{title}{Fighting money laundering with statistics and machine learning},
\newblock \bibinfo{journal}{IEEE Access} \bibinfo{volume}{11} (\bibinfo{year}{2023}) \bibinfo{pages}{8889--8903}.
\bibitem[{{European Union}(2024)}]{EU_AI_Act_2024}
\bibinfo{author}{{European Union}}, \bibinfo{title}{Regulation (eu) 2024/1689 laying down harmonised rules on artificial intelligence (i.e., the artificial intelligence act}, \bibinfo{howpublished}{\url{https://www.europarl.europa.eu/doceo/document/TA-9-2024-0138_EN.pdf}}, \bibinfo{year}{2024}.
\bibitem[{Parente(2024)}]{Parente_2024}
\bibinfo{author}{F.~Parente},
\newblock \bibinfo{title}{The ai act and its impacts on the european financial sector},
\newblock \bibinfo{journal}{The EUROFI Magazine}  (\bibinfo{year}{2024}). \bibinfo{note}{Ghent 2024, DIGITALISATION AND TECHNOLOGY, VIEWS, 128}.
\bibitem[{Xenidis and Senden(2019)}]{xenidis2019eu}
\bibinfo{author}{R.~Xenidis}, \bibinfo{author}{L.~Senden},
\newblock \bibinfo{title}{Eu non-discrimination law in the era of artificial intelligence: Mapping the challenges of algorithmic discrimination},
\newblock \bibinfo{journal}{Rapha{\"e}le Xenidis and Linda Senden,‘EU non-discrimination law in the era of artificial intelligence: Mapping the challenges of algorithmic discrimination’in Ulf Bernitz et al (eds), General Principles of EU law and the EU Digital Order (Kluwer Law International, 2020)}  (\bibinfo{year}{2019}) \bibinfo{pages}{151--182}.
\bibitem[{{European Union Agency for Fundamental Rights} et~al.(2018){European Union Agency for Fundamental Rights}, {European Court of Human Rights}, and {Council of Europe}}]{EUNonDiscriminationLaw2018}
\bibinfo{author}{{European Union Agency for Fundamental Rights}}, \bibinfo{author}{{European Court of Human Rights}}, \bibinfo{author}{{Council of Europe}}, \bibinfo{title}{Handbook on European non-discrimination law}, \bibinfo{edition}{2018} ed., \bibinfo{organization}{European Union Agency for Fundamental Rights}, \bibinfo{address}{Vienna, Austria}, \bibinfo{year}{2018}. \URLprefix \url{https://fra.europa.eu/en/publication/2018/handbook-european-non-discrimination-law-2018-edition}, \bibinfo{note}{accessed: date-of-access}.
\bibitem[{De~Mol(2011)}]{de2011novel}
\bibinfo{author}{M.~De~Mol},
\newblock \bibinfo{title}{The novel approach of the cjeu on the horizontal direct effect of the eu principle of non-discrimination:(unbridled) expansionism of eu law?},
\newblock \bibinfo{journal}{Maastricht journal of European and Comparative law} \bibinfo{volume}{18} (\bibinfo{year}{2011}) \bibinfo{pages}{109--135}.
\bibitem[{Muir(2019)}]{muir2019horizontal}
\bibinfo{author}{E.~Muir},
\newblock \bibinfo{title}{The horizontal effects of charter rights given expression to in eu legislation, from mangold to bauer},
\newblock \bibinfo{journal}{Review of European Administrative Law} \bibinfo{volume}{12} (\bibinfo{year}{2019}) \bibinfo{pages}{185--215}.
\bibitem[{Bertrand et~al.(2021)Bertrand, Maxwell, and Vamparys}]{bertrand2021ai}
\bibinfo{author}{A.~Bertrand}, \bibinfo{author}{W.~Maxwell}, \bibinfo{author}{X.~Vamparys},
\newblock \bibinfo{title}{Do ai-based anti-money laundering (aml) systems violate european fundamental rights?},
\newblock \bibinfo{journal}{International data privacy law} \bibinfo{volume}{11} (\bibinfo{year}{2021}) \bibinfo{pages}{276--293}.
\bibitem[{Sciurba(2019)}]{sciurba2019incompatibility}
\bibinfo{author}{L.~L. Sciurba, Michele}, \bibinfo{title}{The Incompatibility of Global Anti-Money Laundering Regimes with Human and Civil Rights}, \bibinfo{edition}{1} ed., \bibinfo{publisher}{Nomos}, \bibinfo{address}{Baden-Baden}, \bibinfo{year}{2019}.
\bibitem[{{European Union}(2016)}]{GDPR2016}
\bibinfo{author}{{European Union}}, \bibinfo{title}{{General Data Protection Regulation}}, \bibinfo{howpublished}{\url{https://eur-lex.europa.eu/legal-content/EN/TXT/?uri=CELEX\%3A32016R0679}}, \bibinfo{year}{2016}. \bibinfo{note}{Article 5.1(c)}.
\bibitem[{Binns and Veale(2021)}]{binns2021your}
\bibinfo{author}{R.~Binns}, \bibinfo{author}{M.~Veale},
\newblock \bibinfo{title}{Is that your final decision? multi-stage profiling, selective effects, and article 22 of the gdpr},
\newblock \bibinfo{journal}{International Data Privacy Law} \bibinfo{volume}{11} (\bibinfo{year}{2021}) \bibinfo{pages}{319--332}.
\bibitem[{of~Justice of~the European~Union(2023)}]{CJEU_C634_21}
\bibinfo{author}{C.~of~Justice of~the European~Union}, \bibinfo{title}{Judgment of the court of justice of the european union (first chamber) of 7 december 2023: Oq v land hessen, schufa holding ag (case c-634/21)}, \bibinfo{year}{2023}. \URLprefix \url{https://curia.europa.eu/juris/document/document.jsf?mode=DOC&pageIndex=0&docid=280426&part=1&doclang=EN&text=&dir=&occ=first&cid=5526176}, \bibinfo{note}{accessed: 2024-05-23}.
\bibitem[{Shrishak(2025)}]{SHRISHAK2025}
\bibinfo{author}{K.~Shrishak}, \bibinfo{title}{Ai-complex algorithms and effective data protection supervision - effective implementation of data subjects’ rights}, \bibinfo{howpublished}{\url{https://www.edpb.europa.eu/system/files/2025-01/d2-ai-effective-implementation-of-data-subjects-rights_en.pdf}}, \bibinfo{year}{2025}. \bibinfo{note}{Accessed on 6 March 2025}.
\bibitem[{Veale et~al.(2018)Veale, Binns, and Edwards}]{veale2018algorithms}
\bibinfo{author}{M.~Veale}, \bibinfo{author}{R.~Binns}, \bibinfo{author}{L.~Edwards},
\newblock \bibinfo{title}{Algorithms that remember: model inversion attacks and data protection law},
\newblock \bibinfo{journal}{Philosophical Transactions of the Royal Society A: Mathematical, Physical and Engineering Sciences} \bibinfo{volume}{376} (\bibinfo{year}{2018}) \bibinfo{pages}{20180083}.
\bibitem[{Juliussen et~al.(2023)Juliussen, Rui, and Johansen}]{juliussen2023algorithms}
\bibinfo{author}{B.~A. Juliussen}, \bibinfo{author}{J.~P. Rui}, \bibinfo{author}{D.~Johansen},
\newblock \bibinfo{title}{Algorithms that forget: Machine unlearning and the right to erasure},
\newblock \bibinfo{journal}{Computer Law \& Security Review} \bibinfo{volume}{51} (\bibinfo{year}{2023}) \bibinfo{pages}{105885}.
\bibitem[{Leiser and Dechesne(2020)}]{leiser2020governing}
\bibinfo{author}{M.~Leiser}, \bibinfo{author}{F.~Dechesne},
\newblock \bibinfo{title}{Governing machine-learning models: challenging the personal data presumption},
\newblock \bibinfo{journal}{International data privacy law} \bibinfo{volume}{10} (\bibinfo{year}{2020}) \bibinfo{pages}{187--200}.
\bibitem[{King et~al.(2018)King, Walker, and Gurul{\'e}}]{king2018palgrave}
\bibinfo{author}{C.~King}, \bibinfo{author}{C.~Walker}, \bibinfo{author}{J.~Gurul{\'e}}, \bibinfo{title}{The Palgrave handbook of criminal and terrorism financing law}, \bibinfo{publisher}{Springer}, \bibinfo{year}{2018}.
\bibitem[{Commission(2024)}]{ec_aml_cft}
\bibinfo{author}{E.~Commission}, \bibinfo{title}{Anti-money laundering and countering the financing of terrorism at eu level}, \bibinfo{howpublished}{\url{https://finance.ec.europa.eu/financial-crime/anti-money-laundering-and-countering-financing-terrorism-eu-level_en}}, \bibinfo{year}{2024}. \bibinfo{note}{Accessed on 29 May 2024}.
\bibitem[{Kamiran et~al.(2012)Kamiran, Karim, and Zhang}]{kamiran2012decision}
\bibinfo{author}{F.~Kamiran}, \bibinfo{author}{A.~Karim}, \bibinfo{author}{X.~Zhang},
\newblock \bibinfo{title}{Decision theory for discrimination-aware classification},
\newblock in: \bibinfo{booktitle}{2012 IEEE 12th international conference on data mining}, \bibinfo{organization}{IEEE}, \bibinfo{year}{2012}, pp. \bibinfo{pages}{924--929}.
\bibitem[{Kenfack(2023)}]{kenfack2023exploring}
\bibinfo{author}{P.~Kenfack}, \bibinfo{title}{Exploring the landscape of ai ethics}, \bibinfo{howpublished}{\url{https://www.kaggle.com/code/patrikkenfack/exploring-the-landscape-of-ai-ethics/notebook}}, \bibinfo{year}{2023}. \bibinfo{note}{Accessed on 21 May 2024}.
\bibitem[{Wachter et~al.(2020)Wachter, Mittelstadt, and Russell}]{wachter2020bias}
\bibinfo{author}{S.~Wachter}, \bibinfo{author}{B.~Mittelstadt}, \bibinfo{author}{C.~Russell},
\newblock \bibinfo{title}{Bias preservation in machine learning: the legality of fairness metrics under eu non-discrimination law},
\newblock \bibinfo{journal}{W. Va. L. Rev.} \bibinfo{volume}{123} (\bibinfo{year}{2020}) \bibinfo{pages}{735}.
\bibitem[{Quelle(2018)}]{quelle2018enhancing}
\bibinfo{author}{C.~Quelle},
\newblock \bibinfo{title}{Enhancing compliance under the general data protection regulation: the risky upshot of the accountability-and risk-based approach},
\newblock \bibinfo{journal}{European Journal of Risk Regulation} \bibinfo{volume}{9} (\bibinfo{year}{2018}) \bibinfo{pages}{502--526}.
\bibitem[{Shokri et~al.(2017)Shokri, Stronati, Song, and Shmatikov}]{shokri2017membership}
\bibinfo{author}{R.~Shokri}, \bibinfo{author}{M.~Stronati}, \bibinfo{author}{C.~Song}, \bibinfo{author}{V.~Shmatikov},
\newblock \bibinfo{title}{Membership inference attacks against machine learning models},
\newblock in: \bibinfo{booktitle}{2017 IEEE symposium on security and privacy (SP)}, \bibinfo{organization}{IEEE}, \bibinfo{year}{2017}, pp. \bibinfo{pages}{3--18}.
\bibitem[{Jegorova et~al.(2022)Jegorova, Kaul, Mayor, O'Neil, Weir, Murray-Smith, and Tsaftaris}]{jegorova2022survey}
\bibinfo{author}{M.~Jegorova}, \bibinfo{author}{C.~Kaul}, \bibinfo{author}{C.~Mayor}, \bibinfo{author}{A.~Q. O'Neil}, \bibinfo{author}{A.~Weir}, \bibinfo{author}{R.~Murray-Smith}, \bibinfo{author}{S.~A. Tsaftaris},
\newblock \bibinfo{title}{Survey: Leakage and privacy at inference time},
\newblock \bibinfo{journal}{IEEE Transactions on Pattern Analysis and Machine Intelligence}  (\bibinfo{year}{2022}).
\bibitem[{Yeom et~al.(2018)Yeom, Giacomelli, Fredrikson, and Jha}]{yeom2018privacy}
\bibinfo{author}{S.~Yeom}, \bibinfo{author}{I.~Giacomelli}, \bibinfo{author}{M.~Fredrikson}, \bibinfo{author}{S.~Jha},
\newblock \bibinfo{title}{Privacy risk in machine learning: Analyzing the connection to overfitting},
\newblock in: \bibinfo{booktitle}{2018 IEEE 31st computer security foundations symposium (CSF)}, \bibinfo{organization}{IEEE}, \bibinfo{year}{2018}, pp. \bibinfo{pages}{268--282}.
\bibitem[{Choquette-Choo et~al.(2021)Choquette-Choo, Tramer, Carlini, and Papernot}]{pmlr-v139-choquette-choo21a}
\bibinfo{author}{C.~A. Choquette-Choo}, \bibinfo{author}{F.~Tramer}, \bibinfo{author}{N.~Carlini}, \bibinfo{author}{N.~Papernot},
\newblock \bibinfo{title}{Label-only membership inference attacks},
\newblock in: \bibinfo{editor}{M.~Meila}, \bibinfo{editor}{T.~Zhang} (Eds.), \bibinfo{booktitle}{Proceedings of the 38th International Conference on Machine Learning}, volume \bibinfo{volume}{139} of \textit{\bibinfo{series}{Proceedings of Machine Learning Research}}, \bibinfo{publisher}{PMLR}, \bibinfo{year}{2021}, pp. \bibinfo{pages}{1964--1974}. \URLprefix \url{https://proceedings.mlr.press/v139/choquette-choo21a.html}.
\bibitem[{Dibbo et~al.(2023)Dibbo, Chung, and Mehnaz}]{dibbo2023model}
\bibinfo{author}{S.~V. Dibbo}, \bibinfo{author}{D.~L. Chung}, \bibinfo{author}{S.~Mehnaz},
\newblock \bibinfo{title}{Model inversion attack with least information and an in-depth analysis of its disparate vulnerability},
\newblock in: \bibinfo{booktitle}{2023 IEEE Conference on Secure and Trustworthy Machine Learning (SaTML)}, \bibinfo{organization}{IEEE}, \bibinfo{year}{2023}, pp. \bibinfo{pages}{119--135}.
\bibitem[{Kobayashi and Alam(2024)}]{kobayashi2024explainable}
\bibinfo{author}{K.~Kobayashi}, \bibinfo{author}{S.~B. Alam},
\newblock \bibinfo{title}{Explainable, interpretable, and trustworthy ai for an intelligent digital twin: A case study on remaining useful life},
\newblock \bibinfo{journal}{Engineering Applications of Artificial Intelligence} \bibinfo{volume}{129} (\bibinfo{year}{2024}) \bibinfo{pages}{107620}.
\bibitem[{Herm et~al.(2023)Herm, Heinrich, Wanner, and Janiesch}]{herm2023stop}
\bibinfo{author}{L.-V. Herm}, \bibinfo{author}{K.~Heinrich}, \bibinfo{author}{J.~Wanner}, \bibinfo{author}{C.~Janiesch},
\newblock \bibinfo{title}{Stop ordering machine learning algorithms by their explainability! a user-centered investigation of performance and explainability},
\newblock \bibinfo{journal}{International Journal of Information Management} \bibinfo{volume}{69} (\bibinfo{year}{2023}) \bibinfo{pages}{102538}.
\bibitem[{Fern{\'a}ndez et~al.(2018)Fern{\'a}ndez, Garc{\'\i}a, Galar, Prati, Krawczyk, Herrera, Fern{\'a}ndez, Garc{\'\i}a, Galar, Prati et~al.}]{fernandez2018cost}
\bibinfo{author}{A.~Fern{\'a}ndez}, \bibinfo{author}{S.~Garc{\'\i}a}, \bibinfo{author}{M.~Galar}, \bibinfo{author}{R.~C. Prati}, \bibinfo{author}{B.~Krawczyk}, \bibinfo{author}{F.~Herrera}, \bibinfo{author}{A.~Fern{\'a}ndez}, \bibinfo{author}{S.~Garc{\'\i}a}, \bibinfo{author}{M.~Galar}, \bibinfo{author}{R.~C. Prati}, et~al.,
\newblock \bibinfo{title}{Cost-sensitive learning},
\newblock \bibinfo{journal}{Learning from imbalanced data sets}  (\bibinfo{year}{2018}) \bibinfo{pages}{63--78}.
\bibitem[{Hoang and Wiegratz(2023)}]{hoang2023machine}
\bibinfo{author}{D.~Hoang}, \bibinfo{author}{K.~Wiegratz},
\newblock \bibinfo{title}{Machine learning methods in finance: Recent applications and prospects},
\newblock \bibinfo{journal}{European Financial Management} \bibinfo{volume}{29} (\bibinfo{year}{2023}) \bibinfo{pages}{1657--1701}.
\bibitem[{Bockel-Rickermann et~al.(2023)Bockel-Rickermann, Verboven, Verdonck, and Verbeke}]{bockel2023causal}
\bibinfo{author}{C.~Bockel-Rickermann}, \bibinfo{author}{S.~Verboven}, \bibinfo{author}{T.~Verdonck}, \bibinfo{author}{W.~Verbeke},
\newblock \bibinfo{title}{A causal perspective on loan pricing: Investigating the impacts of selection bias on identifying bid-response functions},
\newblock \bibinfo{journal}{arXiv preprint arXiv:2309.03730}  (\bibinfo{year}{2023}).
\bibitem[{Giovanola and Tiribelli(2023)}]{giovanola2023beyond}
\bibinfo{author}{B.~Giovanola}, \bibinfo{author}{S.~Tiribelli},
\newblock \bibinfo{title}{Beyond bias and discrimination: redefining the ai ethics principle of fairness in healthcare machine-learning algorithms},
\newblock \bibinfo{journal}{AI \& society} \bibinfo{volume}{38} (\bibinfo{year}{2023}) \bibinfo{pages}{549--563}.
\bibitem[{Gilbert(2024)}]{gilbert2024eu}
\bibinfo{author}{S.~Gilbert},
\newblock \bibinfo{title}{The eu passes the ai act and its implications for digital medicine are unclear},
\newblock \bibinfo{journal}{npj Digital Medicine} \bibinfo{volume}{7} (\bibinfo{year}{2024}) \bibinfo{pages}{135}.
\bibitem[{Rasheed et~al.(2022)Rasheed, Qayyum, Ghaly, Al-Fuqaha, Razi, and Qadir}]{rasheed2022explainable}
\bibinfo{author}{K.~Rasheed}, \bibinfo{author}{A.~Qayyum}, \bibinfo{author}{M.~Ghaly}, \bibinfo{author}{A.~Al-Fuqaha}, \bibinfo{author}{A.~Razi}, \bibinfo{author}{J.~Qadir},
\newblock \bibinfo{title}{Explainable, trustworthy, and ethical machine learning for healthcare: A survey},
\newblock \bibinfo{journal}{Computers in Biology and Medicine} \bibinfo{volume}{149} (\bibinfo{year}{2022}) \bibinfo{pages}{106043}.

\end{thebibliography}

\end{document}